\documentclass[a4paper,12pt]{article}
\usepackage[margin=2cm]{geometry}
\usepackage{comment}
\usepackage{amssymb,extarrows,graphicx,subfigure,setspace}
\usepackage{cite}
\usepackage{slashed}
\usepackage{tensor}
\usepackage[toc,page]{appendix}
\usepackage{color}
\usepackage{hyperref}
\usepackage{dirtytalk}
\usepackage{titlesec}
\titleformat{\subsubsection}[block]{\color{blue}\Large\bfseries\filcenter}{}{1em}{}
\hypersetup{colorlinks=true, linkcolor=blue, citecolor=red, linktoc=page}
\makeatother

\usepackage{amsmath}
\medskip
\newmuskip\pFqmuskip

\newcommand*\pFq[6][8]{%
  \begingroup 
  \pFqmuskip=#1mu\relax
  \mathcode`\,=\string"8000
  \begingroup\lccode`\~=`\,
  \lowercase{\endgroup\let~}\pFqcomma
  {}_{#2}F_{#3}{\left[\genfrac..{0pt}{}{#4}{#5};#6\right]}%
  \endgroup
}
\newcommand{\pFqcomma}{\mskip\pFqmuskip}

\usepackage{hyperref}

\usepackage{orcidlink}

\newcommand{\be}{\begin{equation}}
\newcommand{\bea}{\begin{eqnarray}}
\newcommand{\eea}{\end{eqnarray}}
\newcommand{\ba}{\begin{array}}
\newcommand{\ea}{\end{array}}
\newcommand{\ee}{\end{equation}}

\numberwithin{equation}{section}
\begin{document}
	\onehalfspacing
	\noindent
	
	\begin{titlepage}
		\vspace{10mm}
		
		
		\vspace*{20mm}
		\begin{center}
			

{\Large {\bf Effect of Non-Extensive Parameter on Page Curve}}

			\vspace*{15mm}
			\vspace*{1mm}
{\bf \large   Ankit Anand\orcidlink{0000-0002-8832-3212}\footnote{\fontsize{8pt}{10pt}\selectfont\ \href{mailto:ankitanandp94@gmail.com}{ankitanandp94@gmail.com}}$^{\dag}$, Dinesh Kumar\footnote{\fontsize{8pt}{10pt}\selectfont\ \href{mailto:kumar.dinesh@galgotiasuniversity.edu.in}{kumar.dinesh@galgotiasuniversity.edu.in}}$^{\star}$ and Aditya Singh\orcidlink{0000-0002-2719-5608}\footnote{\fontsize{8pt}{10pt}\selectfont\ \href{mailto:24pr0148@iitism.ac.in}{24pr0148@iitism.ac.in}\\\;\\ \textbf{\;\;\;\;\;\;\;\;\;\; All Authors Contributed Equally}}$^{\ddag}$} 
  
  \vspace{0.5cm}

{\it $^{\dag}$ Department of Physics, Indian Institute of Technology Kanpur, Kanpur 208016, India\\ 
$^{\star}$Physics Division, School of Basic and Applied Sciences, Galgotias University, Greater Noida 203201, India \\
$^\ddag$ Department of Physics, Indian Institute of Technology (Indian School of Mines), Dhanbad, Jharkhand-826004, India }
			\vspace{0.2cm}

			\vspace*{1cm}

			\vspace{0.2cm}
			\vspace{0.2cm}
		\end{center}
		
\begin{abstract}
This work employs the quantum extremal surface framework to compute the Page curve for black holes corrected by non-extensive entropy. The entropy of Hawking radiation increases linearly with time, leading to the persistence of the information paradox for non-extensive entropy-corrected black holes. At late time, we extremize the generalized entropy functional; incorporating contributions from both matter and the quantum extremal island, we establish that the entanglement entropy of Hawking radiation saturates to the non-extensive extension of the Bekenstein-Hawking entropy. Finally, we study the dependence of non-extensive parameters on the Page time.  

\end{abstract}
\end{titlepage}

\newpage

\section{Introduction}\label{Sec: Introduction}

\quad The study of black hole entropy is crucial for understanding the intricate microscopic structure of spacetime within the framework of quantum gravity. Unlike classical thermodynamic systems, where entropy is proportional to mass and volume, black hole entropy uniquely scales with the horizon area. This deviation from the extensive nature of conventional entropy poses a significant mystery, as the origin of this nonextensivity remains unresolved \cite{Tsallis:2012js}. Building on the foundation of Bekenstein-Hawking entropy, various extensions incorporating nonextensive statistical frameworks have been proposed. Notable examples include Renyi entropy \cite{Renyi} and Tsallis entropy \cite{Tsallis:1987eu}. Recent advancements have introduced alternative formulations such as Barrow entropy \cite{Barrow:2020tzx}, Sharma-Mittal entropy \cite{SayahianJahromi:2018irq}, and Kaniadakis entropy \cite{Kaniadakis:2005zk, Drepanou:2021jiv}. These generalized entropies aim to deepen our understanding of black hole thermodynamics and its connection to quantum gravity. The first law of thermodynamics, which relates entropy, temperature, internal energy, and heat transfer, is sensitive to the choice of entropy definition. Altering the entropy framework, as discussed in \cite{Nojiri:2021czz}, consequently modifies other thermodynamic parameters, offering new perspectives on black hole physics and beyond. Lacking a complete quantum gravity theory, effective field theory (EFT) offers a robust framework for exploring quantum effects near the Planck scale \cite{Donoghue:1994dn}. EFT predicts dynamical non-locality at small distances, modifying black hole metrics to align with quantum-corrected entropy. These corrections, derived via curvature expansions, appear at different orders—third for Schwarzschild \cite{Calmet:2021lny} and second for Reissner-Nordström \cite{Delgado:2022pcc}. Using the Wald entropy formula \cite{Wald:1993nt} and alternative methods \cite{Cano:2019ycn, Yoon:2007aj, Singh:2023hit, Akbar:2003mv, Sadeghi:2014zna, Susskind:1994sm}, quantum-corrected geometries and entropies have been established, deepening our understanding of black hole quantum structure and spacetime dynamics.

\quad The information paradox of black holes has been a central puzzle in theoretical physics ever since the discovery of Hawking radiation \cite{Hawking:1974sw, Hawking:1975vcx, Hawking:1974rv, Hawking:1976ra}. At the semiclassical level, the unitarity of quantum mechanics seems to break down when a black hole is formed from the collapse of a pure quantum state and eventually evaporates, leaving behind only thermal radiation, which is characterized by a mixed quantum state. The key concept is black hole entropy, specifically the Bekenstein-Hawking entropy, which has garnered significant attention over the past few decades proportional to the horizon area; this entropy serves as a cornerstone in the study of quantum gravity \cite{Maldacena:1997de, Solodukhin:2011gn}. However, within the quantum gravity framework, Bekenstein-Hawking entropy is considered a leading-order, semiclassical approximation, with higher-order corrections offering deeper insights. One particularly important refinement is the logarithmic correction to entropy, widely regarded as universal and foundational. This correction emerges from various approaches, including conical singularity and entanglement entropy techniques \cite{Solodukhin:2011gn, Solodukhin:1994yz}, the Euclidean action method \cite{Fursaev:1994te, Sen:2012dw, El-Menoufi:2015cqw, El-Menoufi:2017kew}, conformal anomaly considerations \cite{Cai:2009ua}, the Cardy formula \cite{Carlip:2000nv}, quantum tunneling processes \cite{Banerjee:2008cf, Banerjee:2008fz}, and quantum geometric frameworks \cite{kaul}. The consistent appearance of this logarithmic term across diverse methodologies highlights its critical role in unraveling the microscopic structure of black hole entropy. The Page curve represents the time evolution of this entropy \cite{Page:1993wv, Page:2013dx}. The paradox, therefore, lies in understanding how the Page curve is realized for the entanglement entropy of Hawking radiation. Recently, a potential resolution to this paradox has been proposed through the concept of islands \cite{Penington:2019npb, Almheiri:2019psf, Almheiri:2019hni, Almheiri:2019yqk}. When analyzing the state of Hawking radiation in a region \( R \) outside the black hole, the density matrix of \( R \) is typically obtained by tracing over the states in its complementary region, \(R'\). However, the quantum extremal surface prescription \cite{Ryu:2006bv, Hubeny:2007xt, Engelhardt:2014gca} suggests that certain regions within \(R'\), referred to as islands (\(I \subset R'\)), should not be traced out. As a result, the entanglement entropy of Hawking radiation in \( R \) is effectively determined by considering the states in \( R \cup I \). The entanglement entropy of the Hawking radiation is therefore expressed as:  
\begin{equation}\label{Sgen formula}
 S_{\rm gen} =  \min \left\{\mathrm{ext}\left[\frac{\mathrm{Area}(\partial I)}{4G_{\rm N}}  + S_{\rm matter}(R\cup I) \right]\right\} \ , 
\end{equation}
where the minimal quantum extremal surface prescription is applied. We apply this procedure in the corrected spacetime due to non-extensive generalization. The island rule, initially introduced as part of the conjectured quantum extremal surface prescription, was later derived using the replica method within the framework of the gravitational path integral. When applying the replica trick \cite{Callan:1994py, Holzhey:1994we, Calabrese:2009qy} to gravitational theories, only the boundary conditions of the replica geometries can be fixed. However, new saddle points must also be considered where bulk wormholes connect multiple replicas of spacetime. These configurations, referred to as replica wormholes, naturally lead to the appearance of islands \cite{Penington:2019kki, Almheiri:2019qdq}. In the semi-classical limit, the partition function of the replicated geometry is dominated by the saddle point that minimizes the entanglement entropy. This approach ensures that the replica trick in gravitational theories yields the same formula \eqref{Sgen formula} as the quantum extremal surface prescription. Since replica wormholes are a direct consequence of the replica trick in gravitational systems, the island conjecture is expected to apply universally to any black hole scenario.

\quad The motivation behind this is the study of correction to the black holes from the non-extensive entropy. Studying the information paradox within the framework of non-extensive entropy is particularly worthwhile, as it provides a novel perspective that goes beyond the traditional thermodynamic and statistical mechanics approaches. It has its own right to study the information paradox in non-extensive entropy-corrected black holes and its resolution using the island rule. The structure of the paper is outlined as follows:

\begin{itemize}
    \item \textbf{Section \ref{Sec:Non-Extensive statistic}} provides a concise introduction to the non-extensive generalization of black hole entropy and outlines the methodology employed throughout the study.
    
    \item \textbf{Section \ref{Sec:Barrow's Approach}} delves into Barrow’s approach, deriving the entanglement entropy to first-order corrections in Barrow’s parameter, and subsequently examines the Page curve with and without the inclusion of the island configuration.
    
    \item \textbf{Section \ref{Sec:Kaniadakis's Approach}} explores Kaniadakis’s approach, calculating the entanglement entropy up to first-order corrections in Kaniadakis’s parameter, and evaluates the Page curve under similar conditions.
    
    \item \textbf{Section \ref{Sec:Renyi Approach}} discusses Renyi’s approach, determining the entanglement entropy to first-order corrections in Renyi’s parameter, and investigates the Page curve both with and without the island configuration.
    
    \item \textbf{Section \ref{Sec:Tsallis-Cirto's Approach}} examines the Tsallis-Cirto framework, computing the entanglement entropy up to first-order corrections in Tsallis-Cirto’s parameter, and verifies the Page curve under analogous configurations.
    
    \item \textbf{Section \ref{Sec:Conclusion}} presents the key findings and conclusions of the study.
\end{itemize}

\section{Non-extensive statistic}\label{Sec:Non-Extensive statistic}

The entropy of a black hole is unique in that it scales with its surface area rather than its volume, classifying it as a non-extensive quantity. This implies that black hole entropy is non-additive and adheres to a non-additive composition rule. In contrast, Gibbs thermodynamics assumes entropy to be extensive and additive, meaning it increases proportionally with the size of the system. This assumption relies on neglecting long-range interactions between subsystems, as such forces are considered negligible when the system's size exceeds the range of interactions. Under these conditions, the total entropy is simply the sum of the entropies of its components. However, long-range interactions are significant for systems like black holes and cannot be ignored. As a result, the principles of Gibbs thermodynamics are insufficient for describing black hole thermodynamics. A formulation of non-extensive thermodynamics that is consistent with the zeroth law has remained a long-standing challenge, with a potential solution emerging only recently. To account for the non-extensive behaviour of black hole entropy, several modified frameworks extending Gibbs entropy have been introduced.

  \subsubsection*{I . Barrow Proposal}  

Barrow explores the effects of quantum gravity on the conventional notion of a smooth and uniform event horizon. He introduces a modified entropy formulation for black holes to account for distortions in the event horizon caused by quantum gravity effects. These distortions are quantified using a fractal parameter \( \Delta \). The resulting Barrow entropy \cite{Barrow:2020tzx} is expressed as  
\begin{equation}\label{Barrow Entropy}
	S_B = \left(S_{BH}\right)^{1+\frac{\Delta}{2}} \ ,
\end{equation}
where \( \Delta \) lies in the range \( 0 \leq \Delta \leq 1 \). For \( \Delta = 0 \), the standard Bekenstein-Hawking entropy is recovered, signifying the absence of any fractal structure and \( \Delta = 1 \), signifies the most intricate surface (or maximal deformation) that may account for one higher geometric dimension from an information viewpoint. The above entropy relation (\ref{Barrow Entropy}), associates the black hole area entropy to the horizon area of the black hole. Despite having a completely distinct physical framework, Barrow and Tsallis entropies are comparable. In real terms, Tsallis entropy relies heavily on the equipartition theorem. Regardless of constituting a toy model, we firmly feel that all of these arguments at least consistently support the idea that Barrow's novel understanding of entropy merits more in-depth research.

    \subsubsection*{II. Kaniadakis proposal}

Kaniadakis's proposal \cite{Kaniadakis:2005zk, Kaniadakis:2002zz} arises from a relativistic statistical framework that aligns with the core principles of conventional statistical mechanics. When applied to black hole thermodynamics, Kaniadakis entropy offers a powerful tool for investigating deviations from classical thermodynamic behavior, especially in holographic contexts. The Kaniadakis black hole entropy is expressed as  
\begin{equation}\label{Kaniadakis Entropy}
    S_K = \frac{\sinh(\kappa S_{BH})}{\kappa},  
\end{equation}  
where \( \kappa \) is a deformation parameter that governs the deviation from standard entropy. In the limit \( \kappa \to 0 \), the Kaniadakis entropy smoothly recovers the Bekenstein-Hawking entropy, ensuring consistency with classical results. Its non-extensive nature occurs precisely due to deviations from classical Boltzmann-Gibbs statistics, for systems with long-range interactions or fractal-like structures in particular. As, it exhibits non-extensive thermodynamic behavior and accounts for corrections to quantum gravity, it can be considered as a valuable tool for investigating black hole thermodynamics in modified gravity theories.
\subsubsection*{III. Renyi Proposal}

Renyi entropy generalizes the concept of entropy from information theory and statistical mechanics, offering a robust framework with significant applications in black hole thermodynamics. Unlike the Bekenstein-Hawking entropy, which is extensive and proportional to the event horizon's surface area, Renyi entropy introduces a nonextensive parameter \( \lambda \), enabling a broader description of entropy in systems exhibiting non-standard statistical behavior. By treating \( \lambda \) as a thermodynamic variable, the thermodynamic phase space of black holes has been extended, allowing for the derivation of a modified Smarr relation and an extended first law of black hole thermodynamics. This generalization has revealed that black holes governed by Renyi entropy can exhibit stability properties distinct from those described by traditional Gibbs-Boltzmann statistics \cite{Czinner:2015eyk, Nakarachinda:2022gsb}. In the limit \( \lambda \to 0 \), Renyi entropy reduces smoothly to the Bekenstein-Hawking entropy, ensuring consistency with classical results. The Renyi entropy is given by  
\begin{equation}\label{Renyi Entropy}
    S_R = \frac{\log{\left(1 + \lambda \, S_{BH}\right)}}{\lambda} \ .
\end{equation}

\subsubsection*{IV. Tsallis-Cirto Proposal}
The Tsallis-Cirto entropy is a generalization of the standard Bekenstein-Hawking entropy, incorporating the principles of non-extensive statistical mechanics. This formulation is particularly significant for studying gravitational systems, such as black holes and the universe, where traditional extensive entropy may not be applicable. In cosmology, the Tsallis-Cirto entropy has been utilized to derive modified Friedmann equations that describe the universe's expansion. These modifications offer valuable insights into dark energy and the observed accelerated expansion of the universe. A notable application arises in the framework of entropic cosmology, where gravity is viewed as an emergent phenomenon arising from the statistical behavior of microscopic degrees of freedom. This perspective provides fresh insights into the nature of dark energy and the dynamical evolution of the cosmos \cite{Volovik:2024}. The Tsallis-Cirto entropy can be expressed as
\begin{equation}\label{Tsallis-Cirto Entropy}
    S_{TC} = S_{BH} ^{\;\;\;\;\; \delta+1} \, ,
\end{equation}  
where \( \delta \) is a nonextensive parameter. In the limit \( \delta \to 0 \), the entropy reduces to the standard Bekenstein-Hawking entropy, recovering the classical thermodynamic description.

\quad Now, we assume a spherically symmetric shell of dust with mass \( M \) and an initially large radius. By the Birkhoff theorem, the spacetime outside the shell is described by the Schwarzschild metric, with mass \( M \). Inside the shell, the spacetime remains empty and flat. As the shell collapses, its radius decreases progressively. Once the shell crosses its Schwarzschild radius, a black hole forms. The geometry remains asymptotically flat throughout the process, and the mass \( M \) of the shell, appearing in the Schwarzschild metric, corresponds to the total energy \( E \) of the system, such that \( E = M \). With the help of these assumptions, we derive the temperature related to all the non-extensive generalizations with the help of 
\begin{equation}
    \frac{1}{T_{\text{non-extensive}}} = \left(\frac{dS_{\text{non-extensive}}}{dM}\right)
\end{equation}
and derive the metric function.


\subsubsection*{Methodology}

 The formula \eqref{Sgen formula} consists of two components the first term represents the gravitational contribution to the generalized entropy, denoted as \( S_{\rm gravity} \). This term is proportional to the area (or volume in higher dimensions) of the boundary of the island, \( \partial I \), and is expressed as:
\begin{equation}
S_{\rm gravity} = \frac{{\rm Area}(\partial \mathcal{I})}{4 G_{\rm N}} \ .
\end{equation}
This term reflects the fact that region \( \mathcal{R} \) is separated from the rest of the system, contributing to the entropy of the Hawking radiation. Notably, this term appears even in the case of empty flat spacetime, provided we consider gravitational theories in the bulk. The second term is the matter entanglement entropy \( S_{\rm matter} \) associated with the region \( \mathcal{R} \cup \mathcal{I} \) in the curved spacetime. If no islands are present, the gravitational entropy of an island vanishes. In higher dimensions, it is well-known that the matter entanglement entropy exhibits area-like divergences that depend on the short-distance cutoff \cite{Bombelli:1986rw, Srednicki:1993im}. This leads to a divergence for the matter entropy, given by
\begin{equation}
S_{\rm matter}(\mathcal{R} \cup \mathcal{I}) = \frac{{\rm Area}(\partial \mathcal{I})}{\epsilon^2} + S^{\rm (finite)}_{\rm matter}(\mathcal{R} \cup \mathcal{I})  \ ,
\end{equation}
where \( \epsilon \) represents the short-distance cutoff scale. A similar divergence arises from the boundary of the region \( \mathcal{R} \), but this term is typically neglected in our calculations. This divergence can be absorbed by renormalizing the Newton constant as
\begin{equation}
\frac{1}{4 G^{(r)}_{\rm N}} = \frac{1}{4 G_{\rm N}} + \frac{1}{\epsilon^2} \,,
\end{equation}
where \( G_{\rm N} \) is the bare Newton constant and \( G^{(r)}_{\rm N} \) is the renormalized Newton constant. Given this, if we treat \( G_{\rm N} \) in equation \eqref{Sgen formula} as the renormalized Newton constant, the leading divergence of \( S_{\rm matter}(\mathcal{R} \cup \mathcal{I}) \) is already accounted for. Thus, \( S_{\rm matter}(\mathcal{R} \cup \mathcal{I}) \) contributes only a finite amount, denoted as \( S^{\rm (finite)}_{\rm matter}(\mathcal{R} \cup \mathcal{I}) \). Therefore, in higher dimensions, the revised formula becomes
\begin{equation}
S(R) = \min \left\{\mathrm{ext}\left[ \frac{\mathrm{Area}(\partial \mathcal{I})}{4 G^{(r)}_{\rm N}} + S^{\rm (finite)}_{\rm matter}(\mathcal{R} \cup \mathcal{I}) \right] \right\} \,.
\end{equation}

\begin{figure}
	\begin{center}
		\includegraphics[scale=0.65]{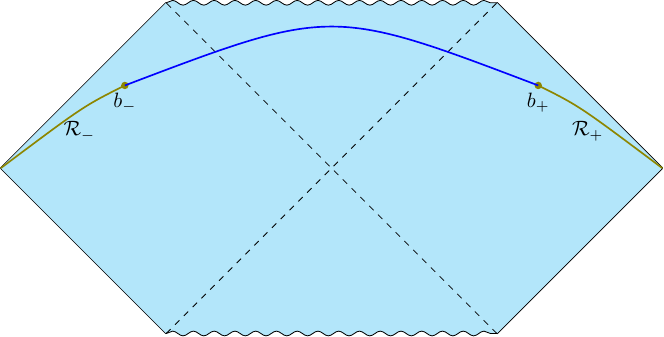}
		\hspace{1cm}
		\includegraphics[scale=0.65]{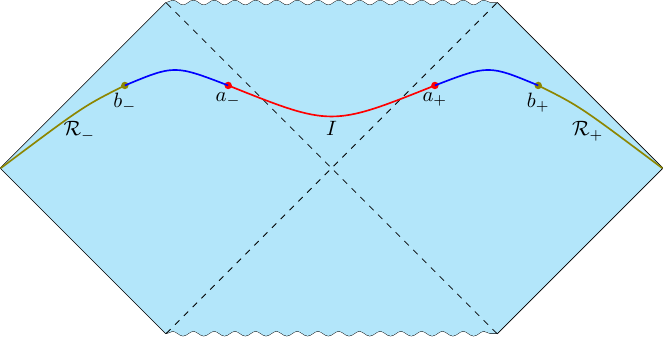}
	\end{center}
	\caption{The Penrose diagram of Schwarchschild black hole. {\it Left:} The configuration without any islands. {\it Right:} The configuration with an island $\mathcal{I}$, shown in red, with endpoints $a_-$ and $a_+$. The radiation region $\mathcal{R} = \mathcal{R}_+ \cup \mathcal{R}_-$ is highlighted in olive, with its endpoints $b_\pm$ located in the baths. Assuming the state on the full Cauchy slice is pure, the entanglement entropy of the matter fields, $S_{\rm matter}(\mathcal{R} \cup \mathcal{I})$, can be calculated. In the left panel, this is evaluated over the complement interval $[b_-, b_+]$, while in the right panel, it is evaluated over $[b_-, a_-] \cup [a_+, b_+]$, as indicated in blue.}
	\label{fig:BH}
\end{figure}
By evaluating this formula using the minimal quantum extremal surface prescription in corrected spacetime, we aim to derive the Page curve. For this, we consider the region for the Hawking radiation \( \mathcal{R} \) in Schwarzschild spacetime. This region consists of two separate parts, \( \mathcal{R}_+ \) and \( \mathcal{R}_- \), located in the right and left wedges of the Penrose diagram, respectively (as illustrated in Fig.~\ref{fig:BH}). As time progresses, the distance between \( \mathcal{R}_+ \) and \( \mathcal{R}_- \) becomes large, and the entanglement entropy of the Hawking radiation without islands grows significantly. Consequently, at late times, the configuration that includes islands is expected to provide the dominant contribution.

To begin, we examine the case without islands. The matter entanglement entropy, in this case, will be calculated for the two separated regions \( \mathcal{R}_+ \) and \( \mathcal{R}_- \). Now, we examine the configuration involving an island \( \mathcal{I} \). At late times, the boundaries of \( \mathcal{I} \) are much closer to the boundary of the region \( \mathcal{R} \) within the same wedge than to the boundaries of \( \mathcal{R}_-\) and \( \mathcal{R}_+\) in the opposite wedge. The volume of an extremal surface connecting these boundaries determines the distance between the boundaries in the right and left wedges\cite{Faulkner:2013ana,Dong:2016hjy}.

\quad To determine the distance, we introduce the Kruskal-like coordinates to achieve the maximal extension of the spacetime\cite{Nutma:2013zea}. These coordinates are defined as:  
\begin{center}
    Right Wedge: \( U = -e^{-\frac{2\pi}{\beta}(t - r_\star)} \) and \( V = e^{\frac{2\pi}{\beta}(t - r_\star)} \) \\ \;\\ 
    Left Wedge: \( U = e^{-\frac{2\pi}{\beta}(t - r_\star)} \) and \( V = -e^{\frac{2\pi}{\beta}(t - r_\star)} \)
\end{center}  
Here, \( \beta \) represents the inverse of non-extensive corrected temperature, and \( r_\star \) is the tortoise coordinate for the non-extensive corrected spacetime. We can easily write the metric in terms of the Kruskal coordinates, which can be represented as 
\begin{equation}\label{Kruskal Metric}
    ds^2 = \Omega_{NE}^{-2} \, dU dV \, ,
\end{equation}  
where \( \Omega_{NE} \) is the conformal factor.

\quad The general formula we will use to calculate the matter contribution to the entanglement entropy is 
\begin{equation}\label{General Matter Entropy}
S_{\text{matter}} = \frac{c}{3} \log \frac{d(a_+, a_-) d(b_+, b_-) d(a_+, b_+) d(a_-, b_-)}{d(a_+, b_-) d(a_-, b_+)} \, , 
\end{equation}  
where \( d(x, y) \) denotes the proper distance between the points \( x \) and \( y \). This formula accounts for the contributions from the boundaries of the entanglement region (\( b_+, b_- \)) and the island boundaries (\( a_+, a_- \)), capturing the interplay between these regions. We analyze the entanglement entropy at early times when no island is present. For the two-dimensional s-wave approximation to hold, the radiation region's points must be sufficiently separated from the sphere's radius. In the absence of an island, the entanglement regions are defined by only two boundary points, located in the right wedge \( \mathcal{R}_+ \) and the left wedge \( \mathcal{R}_- \) (see Fig.~\ref{fig:BH}, Left). At late times, the matter contribution to the entanglement entropy can be expressed as  
\begin{equation}
S^{\text{No Island}}_{\text{matter}} = \frac{c}{3} \log d(b_+, b_-) \ ,
\end{equation}  
where \( b_+ \) and \( b_- \) denote the boundaries of the entanglement regions in the right and left wedges of the non-extensive-corrected geometry. Specifically, \( b_+ \) corresponds to \( (t, r) = (t_b, b) \), and \( b_- \) corresponds to \( (t, r) = (-t_b + i \beta/2, b) \). The imaginary component \( i \beta/2 \) in \( b_- \) arises from its location in the left wedge, introducing a sign reversal for \( U \) and \( V \) in Kruskal coordinates, as detailed below. Using a conformal transformation, the matter contribution to the entanglement entropy in the non-extensive-corrected geometry is determined as  
\begin{equation}\label{Without Island Entropy}
S^{\text{No-Island}}_{\text{matter}} = \frac{c}{6} \log \frac{\left(U(b_-) - U(b_+)\right)\left(V(b_+) - V(b_-)\right)}{\Omega_{NE}(b_+)\Omega_{NE}(b_-)} \ .
\end{equation}  
It is believed that the entropy increases linearly with time and results in a paradox known as the black hole information paradox.

\quad To resolve the entropy paradox at late times, we introduce the island construction and compute the entanglement entropy in the presence of a single island. Since the primary focus is on the late-time behavior, we employ the late-time approximation. At early times, no saddle point exists for the island’s location, implying that the island does not form. For the late-time regime, particularly in the context of the non-extensive-corrected black hole, the radiation entropy becomes significant, as this is the regime where the entropy paradox emerges. We assume that the boundary \( r = b \) of the entanglement region \( \mathcal{R} \) lies far from the black hole's horizon (\( b \gg r_h \)) and use the s-wave approximation to simplify the computations. The total entropy is determined using the matter entropy formula. Using the Kruskal-like coordinates, the generalized entropy \( S_{\rm gen} \) is computed as the sum of the semi-classical fine-grained entropy and the contribution from the area of the quantum extremal surface. The generalized formula for entanglement entropy is 
\begin{eqnarray}\label{With Island Entropy}
    S_{\rm gen} = \text{Contribution from Island's area} + S_{\text{matter}} \ .
\end{eqnarray}

\section{Barrow's Approach}\label{Sec:Barrow's Approach}

In this section, we systematically explore the Barrow-corrected framework by utilizing ~\eqref{Barrow Entropy} to compute the Barrow-corrected temperature, derive the corresponding metric function, and further analyze the spacetime structure. We begin by considering ~\eqref{Barrow Entropy}, which governs the Barrow entropy modification. The Barrow-corrected temperature \( T_B \) can be expressed as  
\begin{equation}
    T_B = \frac{2^{-\Delta-2} \pi ^{-\frac{\Delta}{2}-1} M \left(G M^2\right)^{-\frac{\Delta}{2}-1}}{\Delta+2} \ .
\end{equation}
This corrected temperature accounts for quantum-gravitational effects encoded by the Barrow parameter\cite{DiGennaro:2022grw}. Using the corrected temperature, we construct the Barrow-corrected metric function \( f_B(r) \), which incorporates modifications to the Schwarzschild-like geometry. Assuming a static, spherically symmetric spacetime, the metric function is given as:  
\begin{equation}\label{fB}
    f_B(r) = 1-\frac{2^\Delta (\Delta+2) \pi ^{\Delta/2} G M \left(G M^2\right)^{\Delta/2}}{r} \ ,
\end{equation}  
where \( \Delta \) represents Barrow parameters. By using Eq.~\eqref{fB}, we can find the horizon radius as 
\begin{eqnarray}
    r_h^{{\;\;}\rm Barrow} &=&  \left(4 \pi\right) ^{\Delta /2} G^{1+\Delta/2} M^{1+\Delta} (\Delta +2) \nonumber \\
    &=& 2 G M+\Delta G M \left(\log \left(4 \pi  G M^2\right)+1\right)+O\left(\Delta^2\right) \ . \nonumber
\end{eqnarray}
We can plot this to see how the horizon radius changes with the Barrow parameter. 
\begin{figure}[ht]
    \begin{center}
        \includegraphics[scale=.90]{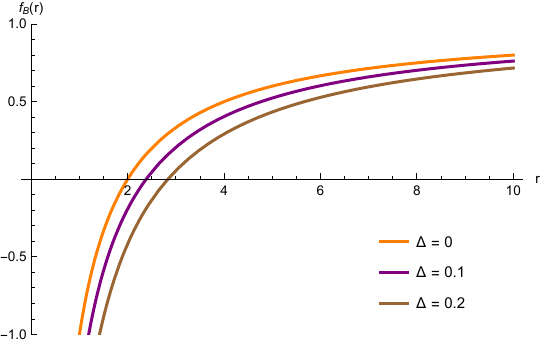} 
    \end{center}
    \caption{Plot of Barrow metric function versus $r$ for various values of Barrow parameters.}
    \label{fig:Barrow metric function}
\end{figure}

We can straightforwardly verify that the metric function\eqref{fB} is a valid solution to the Einstein field equations. Furthermore, the Kretschmann scalar can be explicitly determined as  
\begin{equation}\label{Brrow Krestmann}
\mathcal{K} = R^{\alpha \beta \mu \nu}R_{\alpha \beta \mu \nu}= \frac{3G(\Delta +2)^2 (4 \pi G M^2)^{\Delta +1}}{\pi\, r^6} \ .
\end{equation}
So, in the Barrow-corrected spacetime, the singularity at \( r = r_h^{{\;}\rm Barrow} \) is a coordinate singularity, arising due to the choice of the coordinate and can be resolved through an appropriate transformation. In contrast, the singularity at \( r = 0 \) corresponds to a true curvature divergence and can be seen through the Eq.~\eqref{Brrow Krestmann}, as the Kretschmann scalar in the limit \( r \to 0 \) blows up. 
To remove the co-ordinate singularity, we start with the tortoise coordinate from the metric function, which can be derived as 
\begin{eqnarray}
    r_\star (r) = r+ 2^\Delta (\Delta+2) \pi ^{\Delta/2} G M \left(G M^2\right)^{\Delta/2} \log \left(r-2^\Delta (\Delta+2) \pi ^{\Delta/2} G M \left(G M^2\right)^{\Delta/2}\right) \ .
\end{eqnarray}
This coordinate ensures that the spacetime near the horizon is regular for ingoing and outgoing modes. Now, using Kruskal-like coordinates, the metric can be written similarly to Eq.~\eqref{Kruskal Metric} and the conformal factor \( \Omega_B (r)\) that encapsulates the Barrow corrections and can be expressed as 
\begin{equation}
    \Omega_B (r) = \frac{\beta}{2 \pi }  e^{-\frac{2 \pi  r_\star(r)}{\beta }} \sqrt{1-\frac{2^\Delta (\Delta+2) \pi ^{\Delta/2} G M \left(G M^2\right)^{\Delta/2}}{r}} \ .
\end{equation}  
Applying the conformal mapping technique, the matter's contribution to the entanglement entropy within the Barrow-corrected geometry will be discussed.

\subsection*{Without Island }

Now, we compute the entanglement entropy at early times when no island is present. The radiation region points must be sufficiently separated for the two-dimensional s-wave approximation to be valid. Applying a conformal transformation and using Eq.~\eqref{Without Island Entropy}, we find that the matter contribution to the entanglement entropy in Barrow-corrected geometry is
\begin{eqnarray}\label{S without Island}
     S_{\rm gen.} &=& \frac{c}{6}\log{\left[\frac{  \beta \left(b-2^{\Delta } \pi ^{\Delta /2} (\Delta +2) G M \left(G M^2\right)^{\Delta /2}\right)}{\pi b}\cosh ^2\left(\frac{2 \pi  t_b}{2^{\Delta+2} (\Delta+2) \pi ^{\Delta/2+1} G M \left(G M^2\right)^{\Delta/2} }\right)\right]} \nonumber \\
      & \approx & \frac{c}{6} \left(\frac{2 \pi  t_b}{2^{\Delta+2} (\Delta+2) \pi ^{\Delta/2+1} G M \left(G M^2\right)^{\Delta/2} }\right) \nonumber \\
      & \approx& \frac{c}{24 G M} \; t_b-\Delta \frac{c \left(\log \left(G M^2\right)+1+2 \log (2)+\log (\pi )\right)}{48 (G M)} \; t_b + \mathcal{O}\left(\Delta ^2\right) \ .
 \end{eqnarray}

Without the island, the information remains trapped within the black hole, causing the entanglement entropy to increase linearly over time. In this scenario, no Page time emerges, and the radiation entropy eventually becomes infinitely larger than the Barrow entropy of the black hole this leads to a paradox, which we will address in the following section, where we show that the island construction resolves this issue and yields the correct Page curve for the Barrow-corrected black holes\cite{Anand:2024txo}.

\subsection*{With island}

We will now introduce the concept of island construction and compute the entanglement entropy in the presence of a single island. For the Barrow-corrected black hole, our focus is on the large-time limit of the entanglement entropy—specifically, the radiation at late times—since this is when entropy becomes problematic. Again, using the Kruskal-like coordinates, the generalized entropy is calculated as the semi-classical fine-grained entropy plus the contribution from the area of the quantum extremal surface as in Eq.~\eqref{With Island Entropy}. Since we are interested in the behavior of the system at late times, we adopt the late-time approximation because, at early time, there is no saddle point that exists for the island's location, which implies that the island does not form and the entropy is the same as we discuss without the island. We have used the approximations  
\begin{eqnarray}
    \frac{1}{2} \sqrt{\frac{b-2^{\Delta } \pi ^{\Delta /2} (\Delta +2) G_N M \left(G_N M^2\right)^{\Delta /2}}{a-2^{\Delta } \pi ^{\Delta /2} (\Delta +2) G_N M \left(G_N M^2\right)^{\Delta /2}}}\; e^{\left(\frac{ M (b-a) \left(G_N M^2\right)^{-\frac{\Delta }{2}-1}}{(\Delta +2)2^{\Delta +1} \pi ^{\frac{\Delta }{2}}}\right)} &\ll& \cosh \left(\frac{ (t_a+t_b) \left(G_N M^2\right)^{-\frac{\Delta }{2}}}{(\Delta +2) G_N M \pi ^{\frac{\Delta }{2}}2^{\Delta +1} }\right) \nonumber \\
    \frac{1}{2} \sqrt{\frac{b-2^{\Delta } \pi ^{\Delta /2} (\Delta +2) G_N M \left(G_N M^2\right)^{\Delta /2}}{a-2^{\Delta } \pi ^{\Delta /2} (\Delta +2) G_N M \left(G_N M^2\right)^{\Delta /2}}} \; e^{\left(\frac{ M (b-a) \left(G_N M^2\right)^{-\frac{\Delta }{2}-1}}{(\Delta +2)2^{\Delta +1} \pi ^{\frac{\Delta }{2}}}\right)} &\gg& \cosh \left(\frac{ (t_a-t_b) \left(G_N M^2\right)^{-\frac{\Delta }{2}}}{(\Delta +2) G_N M \pi ^{\frac{\Delta }{2}}2^{\Delta +1} }\right)  \ . \nonumber
\end{eqnarray}
The generalized entropy $S_{\rm gen}$ is
\begin{eqnarray}
    S_{\rm gen} &=& \frac{2 \pi  a^2}{G_N}+\frac{1}{6} c \log \Bigg(\frac{2^{4 \Delta +4} \pi ^{2 \Delta } (\Delta +2)^4 G_N^4 M^4 \left(G_N M^2\right)^{2 \Delta } \left(b-2^{\Delta } \pi ^{\Delta /2} (\Delta +2) G_N M \left(G_N M^2\right)^{\Delta /2}\right)^2 }{a b}\Bigg)\nonumber \\
    && -\frac{2}{3} c \sqrt{\frac{a-2^{\Delta } \pi ^{\Delta /2} (\Delta +2) G_N M \left(G_N M^2\right)^{\Delta /2}}{b-2^{\Delta } \pi ^{\Delta /2} (\Delta +2) G_N M \left(G_N M^2\right)^{\Delta /2}}} \exp \left(\frac{2^{-\Delta -1} \pi ^{-\frac{\Delta }{2}} (a-b) \left(G_N M^2\right)^{-\frac{\Delta }{2}}}{(\Delta +2) G_N M}\right) \nonumber \\
    &&\cosh \left(\frac{2^{-\Delta -1} \pi ^{-\frac{\Delta }{2}} (t_a-t_b) \left(G_N M^2\right)^{-\frac{\Delta }{2}}}{(\Delta +2) G_N M}\right)+\frac{1}{6} c \log \Bigg(\frac{\exp \left(\frac{2^{-\Delta } \pi ^{-\frac{\Delta }{2}} (b-a) \left(G_N M^2\right)^{-\frac{\Delta }{2}}}{(\Delta +2) G_N M}\right)}{ab}\Bigg) \ .
\end{eqnarray}
By first extremizing the given expression with respect to \(t_a\), the only solution obtained is \(t_a = t_b\). Consequently, the focus shifts to extremizing with respect to \(a\), which determines the location of the island as
\begin{eqnarray}
   a &=& 2^{\Delta } \pi ^{\Delta /2} (\Delta +2) G_N M \left(G_N M^2\right)^{\Delta /2} \nonumber \\
  &&  -\frac{c^2 (4 \pi )^{\Delta } (\Delta +2)^2 G_N \left(G_N M^2\right)^{\Delta +1} \exp \left(1-\frac{b 2^{-\Delta } \pi ^{-\frac{\Delta }{2}} M \left(G_N M^2\right)^{-\frac{\Delta }{2}-1}}{\Delta +2}\right)}{\left(2^{\Delta } \pi ^{\Delta /2} (\Delta +2) G_N M \left(G_N M^2\right)^{\Delta /2}-b\right) \left(c-3 (4 \pi )^{\Delta +1} (\Delta +2)^2 \left(G_N M^2\right)^{\Delta +1}\right)^2} \ .
\end{eqnarray}
By incorporating the island's position, the entanglement entropy can be expressed as
\begin{eqnarray}\label{S with Island}
    S_{\rm gen}&=&\frac{\Delta  \left(\log \left(4 \pi  G_N M^2\right)+1\right) \left(3 b^2-3 b G_N M \left(c+32 \pi  G_N M^2+2\right)+2 G_N^2 M^2 \left(5 c+96 \pi  G_N M^2\right)\right)}{12 G_N M (2 G_N M-b)} \nonumber \\
    && + \left(\frac{1}{6} c \log \left(\frac{128 G_N^3 M^3 (b-2 G_N M)^2}{b}\right)+\frac{b}{2 G_N M}+8 \pi  G_N M^2-1\right) + \mathcal{O}(\Delta^2) \ .
\end{eqnarray}
This result demonstrates that the entropy remains constant over time, providing a resolution to the entropy paradox in the late-time regime.
\begin{figure}[ht]
    \begin{center}
        \includegraphics[scale=0.75]{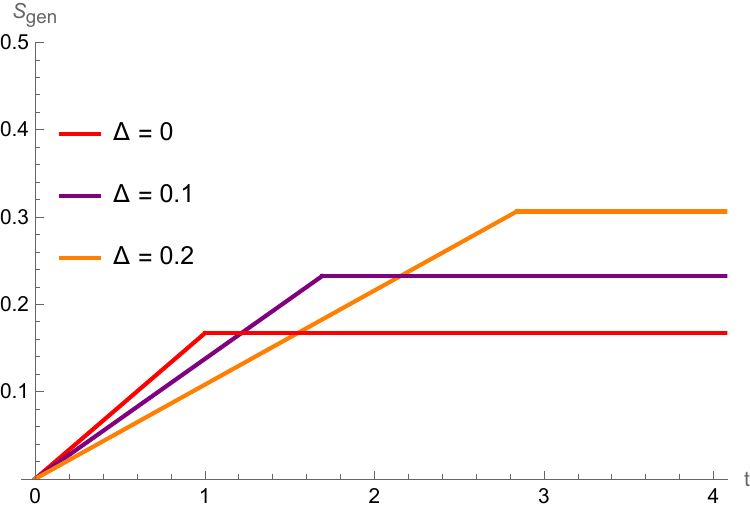}
    \end{center}
    \caption{Plot of $S_{\rm gen}$ with and without island.}
    \label{fig:Barrow Sgen}
\end{figure}
In conclusion, the analysis of the result \eqref{S without Island} and the result \eqref{S with Island} establishes that the island configuration manifests at late times, effectively terminating the growth of entanglement entropy. The island's boundary is located close to the event horizon\cite{Jusufi:2021fek, Abreu:2024tdv}, contributing precisely to the Barrow entropy associated with the Barrow corrected black hole.
\subsection{Page time}

The Page time refers to the moment when the entropy of radiation reaches its maximum. For an evaporating black hole, the point after which the entropy of the radiation begins to decrease occurs approximately when the black hole has lost half of its initial mass. The entropy remains constant after the Page time for an eternal black hole. The Page time for a Barrow-corrected black hole can be determined by comparing the expressions for the entropy without an island \eqref{S without Island} and with an island \eqref{S with Island}. The transition from the no-island configuration to the island configuration occurs approximately at the point where these two entropy curves intersect. After this transition, the entanglement entropy remains constant. Using this reasoning, the Page time is
\begin{equation}
t_{\text{Page}} \sim \frac{96 \pi ^3 G_N^3 M^3}{c}+\frac{144 \pi ^3 \Delta  G_N^3 M^3 \left(\log \left(4 \pi  G_N M^2\right)+1\right)}{c}+ \mathcal{O}\left(\Delta ^2\right) \ .
\end{equation}
By combining the results from Eq.~\eqref{S without Island} and Eq.~\eqref{S with Island}, the Page curve for the evolution of radiation entropy (or equivalently the entropy of the black hole) can be derived, as shown in Fig.\ref{fig:Barrow Sgen}. Initially, the entropy of the radiation increases almost linearly with time, as no island is formed. Around the Page time, an island forms near the black hole’s horizon, causing the entropy to stabilize at a nearly constant value.

\section{Kaniadakis's Approach}\label{Sec:Kaniadakis's Approach}  

\begin{figure}[ht]
    \begin{center}
        \includegraphics[scale=.90]{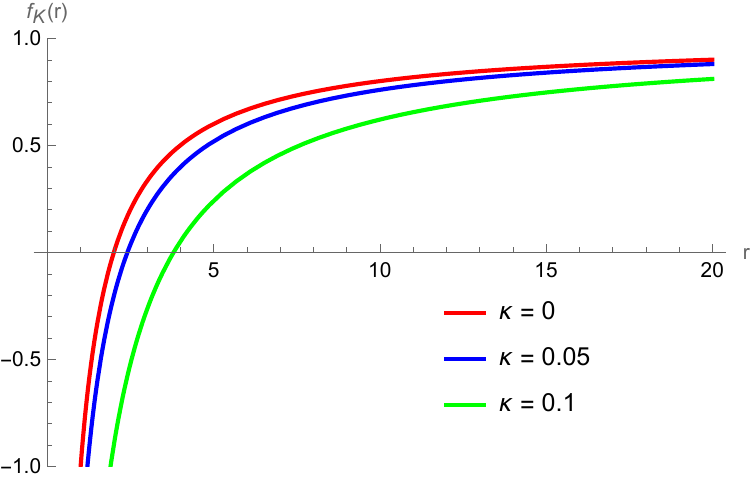} 
    \end{center}
    \caption{Plot of the Kaniadakis metric function versus \( r \) for different values of the Kaniadakis parameter.}
    \label{fig:Kaniadakis metric function}
\end{figure}

In this section, we systematically investigate the Kaniadakis-corrected framework by utilizing Eq.~\eqref{Kaniadakis Entropy} to derive the Kaniadakis-corrected temperature, construct the corresponding metric function, and analyze the modified spacetime structure. We begin by considering the Kaniadakis entropy modification governed by Eq.~\eqref{Kaniadakis Entropy}. The Kaniadakis-corrected temperature \( T_K \) is given by
\begin{equation}
T_K =\frac{\text{sech}\left(4 \pi  G \kappa  M^2\right)}{8 \pi  G M} \ ,
\end{equation}  
which incorporates quantum gravitational effects as characterized by the Kaniadakis parameter \( \kappa \). This corrected temperature represents the deviations from the standard Schwarzschild temperature, influenced by the quantum-gravitational modifications to the spacetime geometry\cite{Lewkowycz:2013nqa, Nojiri:2022dkr}. Using this corrected temperature, we proceed to construct the Kaniadakis-corrected metric function \( f_K(r) \), which modifies the Schwarzschild geometry to account for the Kaniadakis parameter. Assuming a static, spherically symmetric spacetime, the metric function is expressed as:
\begin{equation}\label{fK}
f_K(r) = 1-\frac{2 G M \cosh \left(4 \pi  \kappa\; G  M^2\right)}{r} \ .
\end{equation}  
Here, \( \kappa \) represents the Kaniadakis parameter, quantifying the deviation from the Schwarzschild black hole. The metric function \( f_K(r) \) reflects the changes in the horizon radius due to the quantum-gravitational modifications. A plot of the Kaniadakis metric function as a function of \( r \) for various values of \( \kappa \) is shown in Fig.\ref{fig:Kaniadakis metric function}.
Utilizing Eq.~\eqref{fK}, the horizon radius can be determined as  
\begin{eqnarray}  
r_h^{\;\;\rm Kaniadakis} &=&  2 G M \cosh \left(4 \pi  \kappa  G M^2\right) \nonumber \\
&=& 2 G M+16 \pi ^2 G^3 \kappa ^2 M^5+ \mathcal{O}\left(\kappa ^4\right) \ . \nonumber
\end{eqnarray}  
Furthermore, a direct verification confirms that the metric function \eqref{fK} satisfies the Einstein field equations. The Kretschmann scalar, which characterizes the curvature properties of spacetime, is explicitly given by  
\begin{equation} \label{Kaniadakis Krestmann}  
\mathcal{K} = \frac{48 G^2 M^2 \cosh ^2\left(4 \pi  \kappa G M^2\right)}{r^6} \ .
\end{equation}  
Again, similar to the Barrow-corrected case, \( r = r_h^{\;\rm Kaniadakis} \) is a coordinate singularity, and \( r = 0 \) represents a true curvature singularity, as evidenced by Eq.~ \eqref{Kaniadakis Krestmann}. To eliminate the coordinate singularity, we introduce the tortoise coordinate, derived from the metric function, as follows
\begin{equation}
r_\star (r) = r + 2 G M \cosh \left(4 \pi  G \kappa  M^2\right) \log \left(r-2 G M \cosh \left(4 \pi  G \kappa  M^2\right)\right) \ .
\end{equation}  
The metric in these coordinates can be written similarly to Eq.~\eqref{Kruskal Metric} with identifying \( \Omega_K(r) \) as the conformal factor that incorporates the Kaniadakis corrections. The conformal factor \( \Omega_K (r) \) is given by
\begin{equation}
\Omega_K (r) = \frac{\beta}{2 \pi }  \exp{\left(-\frac{2 \pi  r_\star(r)}{\beta }\right)} \sqrt{1-\frac{2 G M \cosh \left(4 \pi  G \kappa  M^2\right)}{r}}  \ .
\end{equation}  
This factor encodes the quantum modifications to the geometry, driven by the Kaniadakis parameter. By employing Kruskal-like coordinates and this conformal factor, we obtain a description of the Kaniadakis-corrected spacetime in these coordinates. Again, we apply the conformal mapping technique to investigate the matter's contribution to the entanglement entropy within the Kaniadakis-corrected geometry. 

\subsection*{Without Island}

Since no island forms during the early stages of the evaporation process. We use the Eq.~\eqref{Without Island Entropy} and compute the matter contribution to the entanglement entropy in Kandiakis-corrected geometry as 
\begin{equation}
S_{\rm gen.} = \frac{c}{6} \log \left[\frac{\beta \cosh \left(\frac{2 \pi  t_b}{\beta }\right)}{\pi }\frac{ \left(b-2 G M \cosh \left(4 \pi  G \kappa  M^2\right)\right)}{b}\right] \ ,
\end{equation}  
which simplifies the approximation 
\begin{equation}\label{S without Island K}
S_{\rm gen.} \approx \frac{\pi  c t_b}{3} + \frac{c \log\left(b-2 G M \cosh \left(4 \pi  G \kappa  M^2\right)\right)}{6 \pi ^2 b} \ .
\end{equation}  
\quad In the absence of an island, the entanglement entropy continues to grow linearly with time as the information remains trapped inside the black hole. This linear growth leads to the entanglement entropy eventually surpassing the Kaniadakis-corrected entropy of the black hole, creating a paradox. The radiation entropy becomes infinitely larger than theKadiakis entropy. To resolve this issue, the introduction of the island construction is necessary.

\subsection*{With Island}

To address the entropy paradox at late times, we now incorporate the island construction and compute the entanglement entropy in the presence of a single island. We adopt the s-wave approximation to simplify calculations, and using the matter entropy formula, the total entropy is determined by using Kruskal-like coordinates, and the generalized entropy \( S_{\rm gen} \) can be computed with the help of Eq.~\eqref{With Island Entropy} and by using approximations 
\begin{eqnarray}
    t_a + t_b &\gg& 4 G_N M \cosh \left(4 \pi  G_N \kappa  M^2\right) \cosh^{-1}{\left(\frac{1}{2} \sqrt{\frac{b-2 G_N M \cosh \left(4 \pi  G_N \kappa  M^2\right)}{a-2 G_N M \cosh \left(4 \pi  G_N \kappa  M^2\right)}} e^{\frac{(b-a) \text{sech}\left(4 \pi  G_N \kappa  M^2\right)}{4 G_N M}}\right)} \nonumber \\
    t_a-t_b &\ll& 4 G_N M \cosh \left(4 \pi  G_N \kappa  M^2\right) \cosh^{-1}{\left(\frac{1}{2} \sqrt{\frac{b-2 G_N M \cosh \left(4 \pi  G_N \kappa  M^2\right)}{a-2 G_N M \cosh \left(4 \pi  G_N \kappa  M^2\right)}} e^{\frac{(b-a) \text{sech}\left(4 \pi  G_N \kappa  M^2\right)}{4 G_N M}}\right)} \ . \nonumber 
\end{eqnarray}
The generalized entropy \( S_{\rm gen} \) is 
\begin{equation}
\begin{aligned}
    S_{\rm gen} &= \frac{2 \pi  a^2}{G_N}-\frac{2}{3} c \sqrt{\frac{a-2 G_N M \cosh \left(4 \pi  G_N \kappa  M^2\right)}{b-2 G_N M \cosh \left(4 \pi  G_N \kappa  M^2\right)}} e^{\frac{(a-b) \text{sech}\left(4 \pi  G_N \kappa  M^2\right)}{4 G_N M}} \cosh \left(\frac{(t_a-t_b) \text{sech}\left(4 \pi  G_N \kappa  M^2\right)}{4 G_N M}\right) \\
    & +\frac{1}{6} c \log \left(\frac{256 G_N^4 M^4 \cosh ^4\left(4 \pi  G_N \kappa  M^2\right) \left(b-2 G_N M \cosh \left(4 \pi  G_N \kappa  M^2\right)\right)^2 e^{\frac{(b-a) \text{sech}\left(4 \pi  G_N \kappa  M^2\right)}{2 G_N M}}}{a b}\right) \ .
\end{aligned}
\end{equation}

By first extremizing this expression with respect to \( t_a \), the only solution obtained is \( t_a = t_b \). Subsequently, the focus shifts to extremizing with respect to \( a \), which determines the location of the island as
\begin{eqnarray}
   a = 2 G_N M + \frac{4 G_N^2 M^2 e^{4 \pi ^2 b G_N \kappa ^2 M^3-\frac{b}{2 G_N M}+1}}{b} \ .
\end{eqnarray}
By substituting the location of the island, we have the entanglement entropy as 
\begin{eqnarray}\label{S with Island K}
    S_{\rm gen} &=& \left(\frac{1}{6} c \log \left(\frac{128 G_N^3 M^3 (b-2 G_N M)^2}{b}\right)+\frac{b}{2 G_N M}+8 \pi  G_N M^2-1\right) \\
    && -\frac{4 \kappa ^2 \left(\pi ^2 G_N M^3 \left(3 b^2-3 b G_N M \left(c+32 \pi  G_N M^2+2\right)+2 G_N^2 M^2 \left(5 c+96 \pi  G_N M^2\right)\right)\right)}{3 (b-2 G_N M)} \ . \nonumber 
\end{eqnarray}
This result indicates that the entropy does not grow with time, confirming the resolution of the entropy paradox at late times. Combining the results for both early and late times, we obtain the Page curve,
\begin{figure}[ht]
    \begin{center}
        \includegraphics[scale=0.75]{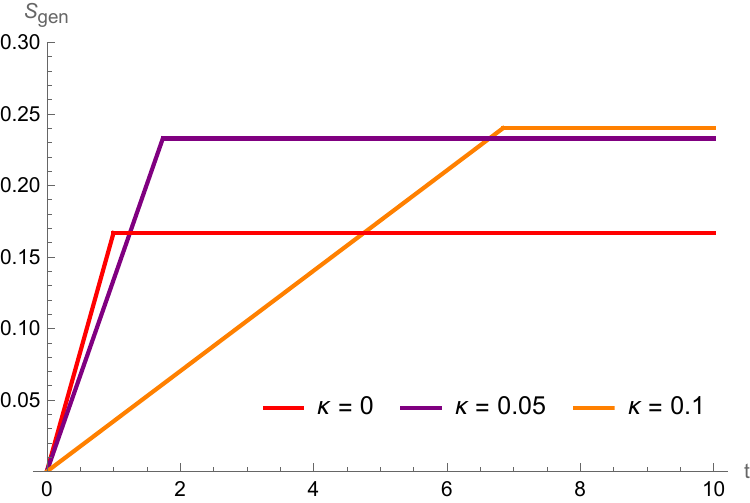}
    \end{center}
    \caption{Plot of $S_{\rm gen}$ with and without island.}
    \label{fig:Kandiakis Sgen}
\end{figure}
which encapsulates the initial linear growth of entropy in Eq.~\eqref{S without Island K} followed by its eventual saturation to a constant value in Eq.~\eqref{S with Island K}. This behavior demonstrates the preservation of information during black hole evaporation and resolves the information paradox. 
\subsection*{Page Time}

In the Kandiakis-corrected framework, the Page time can be determined by comparing the entropy expressions for configurations without an island \eqref{S without Island K} and with an island \eqref{S with Island K}. The transition from a no-island to an island arrangement is represented by the intersection of these two entropy curves, which is precisely the Page time. The entanglement entropy stabilizes at an almost constant value following this transition. The Kandiakis-corrected Page time is given by
\begin{equation}
t_{\text{Page}} \sim \frac{96 \pi ^3 G^3 M^3}{c}+\frac{2304 \pi ^5 G^5 \kappa ^2 M^7}{c} 
\end{equation}  
where \( \kappa \) is the Kandiakis correction parameter, \( G_N \) is the Newton constant, \( M \) is the black hole mass, and \( c \) is a proportionality constant. This result encapsulates the influence of Kandiakis's corrections on entanglement entropy dynamics. By combining the entropy expressions from \eqref{S without Island K} and \eqref{S with Island K}, the Page curve illustrating the evolution of radiation entropy (or equivalently, black hole entropy) can be derived. At first, without an island, the radiation entropy grows linearly with time. The entropy stabilizes around the Page time when an island forms close to the black hole horizon. The Page curve for the Kandiakis-corrected black hole entropy is shown in Fig.~\ref{fig:Kandiakis Sgen}.

\section{Renyi's Approach}\label{Sec:Renyi Approach}

In this section, we systematically investigate the Renyi-corrected framework by utilizing Eq.~\eqref{Renyi Entropy} to derive the Renyi-corrected temperature, compute the associated metric function, and analyze the spacetime structure. Beginning with Eq.~\eqref{Renyi Entropy}, which defines the modification introduced by the Renyi entropy, we derive the corrected temperature \( T_R \), expressed as
\begin{equation}
    T_R = \frac{1}{8 \pi  G M}+\frac{\lambda  M}{2} \ .
\end{equation}  
This temperature captures quantum-gravitational effects encoded by the Renyi parameter \( \lambda \). Using this result, we construct the Renyi-corrected metric function \( f_R(r) \), which reflects deviations from the standard Schwarzschild geometry. For a static, spherically symmetric spacetime, the metric function is given by 
\begin{equation}\label{fR}
    f_R(r) = 1-\frac{2 G M}{4 \pi  G \lambda  M^2 r+r} \ ,
\end{equation}  
where \( \lambda \) quantifies the Renyi entropy corrections. A plot of this function reveals the dependence of the horizon radius on the Renyi parameter. 
\begin{figure}[ht]
    \begin{center}
        \includegraphics[scale=.90]{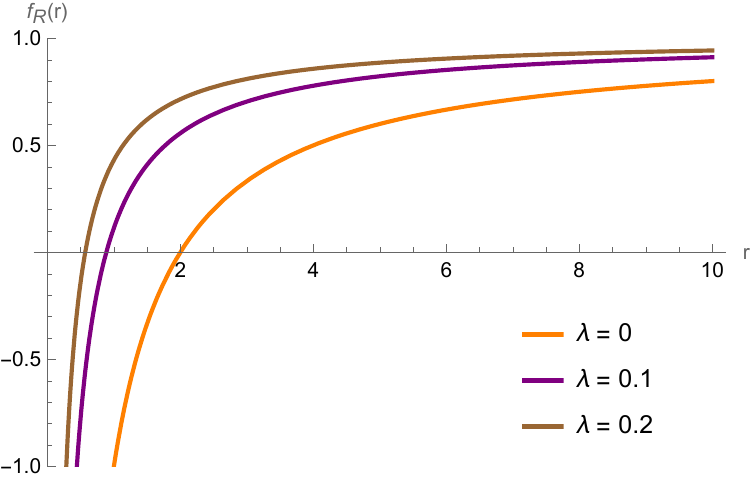} 
    \end{center}
    \caption{Plot of the Renyi metric function versus \( r \) for various values of the Renyi parameter.}
    \label{fig:Renyi metric function}
\end{figure}
By employing Eq.~\eqref{fR}, the horizon radius in the Renyi-corrected spacetime is expressed as
\begin{eqnarray}
r_h^{{\;} \rm Renyi} &=&  \frac{2 G M}{4 \pi  \lambda  G M^2+1} \nonumber \\ 
    &=& 2 G M-8 \lambda  \pi  G^2 M^3 + \mathcal{O}\left(\lambda ^2\right) \ . \nonumber
\end{eqnarray}  
The correction comes due to the correction parameter $\lambda$ in the horizon radius. A straightforward calculation confirms that the metric function \eqref{fR} is indeed a valid solution to the Einstein field equations. Additionally, the Kretschmann scalar, which serves as an invariant measure of spacetime curvature, is explicitly computed as  
\begin{equation} \label{Renyi Krestmann}  
\mathcal{K}  = \frac{48 G^2 M^2}{r^6 \left(4 \pi  \lambda  G M^2+1\right)^2} \ . 
\end{equation}  
The singularity at \( r = r_h^{{\;}\rm Renyi} \) is a coordinate singularity, which emerges due to the specific choice of coordinates, and the singularity at \( r = 0 \) is a real curvature singularity, as evidenced by the divergent behavior of the Kretschmann scalar, shown in Eq.~ \eqref{Renyi Krestmann}, which diverges as \( r \to 0 \), indicating an intrinsic breakdown of spacetime geometry. To resolve the coordinate singularity, we introduce the tortoise coordinate, derived from the metric function, as follows 
\begin{eqnarray}
    r_\star(r) = \frac{2 G M (2 \pi  \lambda  M r-1)+2 G M \log (2 G M (2 \pi  \lambda  M r-1)+r)+r}{4 \pi  G \lambda  M^2+1} \ .
\end{eqnarray}  
Using the Kruskal-like coordinates, the metric takes the form of Eq.~\eqref{Kruskal Metric} and the conformal factor \( \Omega_R(r) \) incorporating Renyi corrections, given by
\begin{equation}
    \Omega_R(r) = \frac{\beta  e^{-\frac{2 \pi  r_\star(r)}{\beta }} }{2 \pi } \sqrt{1-\frac{2 G M}{4 \pi  G \lambda  M^2 r+r}} \ .
\end{equation}  
Following the conformal mapping approach, we discuss the implications of these modifications for the matter’s contribution to the entanglement entropy within the Renyi-corrected geometry.  
\subsection*{Without Island}  

When no island forms during the initial stage of the evaporation process, the entanglement entropy grows linearly and can be verified by using Eq.~\eqref{Without Island Entropy}. In the Kruskal-like coordinates, the matter contribution to the entanglement entropy in the Renyi-corrected geometry becomes 
\begin{equation}
S_{\rm gen.} = \frac{\beta }{\pi} \cosh \left(\frac{2 \pi  t}{\beta }\right) \sqrt{\frac{ \left(4 \pi  b G \lambda  M^2+b-2 G M\right)}{4 \pi  b G \lambda  M^2+b}} \ ,
\end{equation}  
which simplifies under the large \( t_b \) approximation to
\begin{equation}\label{Renyi without island}
S_{\rm gen.} \approx \frac{c t_b}{6}+\frac{2}{3} \pi  c \lambda  t_b \ .
\end{equation}  

In the absence of an island, the entanglement entropy grows linearly with time, is verified, and results in the information remaining confined within the black hole. This linear growth ultimately leads the entropy to exceed the Renyi-corrected black hole entropy, creating a paradox. Without the emergence of a Page time, the radiation entropy grows infinitely larger than the black hole’s entropy, violating unitarity and information conservation. This highlights the necessity of introducing the island construction to resolve the paradox and recover a consistent Page curve.

\subsection*{With Island}

We now introduce the concept of island construction to compute the entanglement entropy in the presence of a single island. For the Renyi-corrected black hole, we focus on the large-time behavior of the entanglement entropy, particularly the radiation entropy at late times, where the growth of entropy poses significant challenges. Using Eq.~\eqref{General Matter Entropy}, we can compute the entanglement entropy for conformal matter as  

\begin{eqnarray}
S_{\text{matter}} &=& \frac{c}{6}\log \left(\frac{256 G_N^4 M^4 \left(4 \pi  b G_N \lambda  M^2+b-2 G_N M\right)^2 e^{\frac{(b-a) \left(4 \pi  G_N \lambda  M^2+1\right)}{2 G_N M}}}{a b \left(4 \pi  G_N \lambda  M^2+1\right)^6}\right) \nonumber \\
&& -\frac{2c}{3} \left(e^{\frac{(a-b) \left(4 \pi  G_N \lambda  M^2+1\right)}{4 G_N M}} \sqrt{\frac{4 \pi  a G_N \lambda  M^2+a-2 G_N M}{4 \pi  b G_N \lambda  M^2+b-2 G_N M}} \cosh \left(\frac{(t_a-t_b) \left(4 \pi  G_N \lambda  M^2+1\right)}{4 G_N M}\right)\right) \ . \nonumber
\end{eqnarray}  
Now, we can compute the generalized entropy \( S_{\rm gen} \) by using Eq.~\eqref{With Island Entropy} and using the approximations as 
\begin{eqnarray}
    \cosh \left(\frac{t_a+t_b}{4 G_N M}\right)+\pi  \lambda  M (t_a+t_b) \sinh \left(\frac{t_a+t_b}{4 G_N M}\right) &\gg& \Bigg\{\frac{1}{2} e^{\frac{b-a}{4 G_N M}} \sqrt{\frac{b-2 G_N M}{a-2 G_N M}} \nonumber \\
    && -\frac{\lambda  \left(\pi  M (a-b) e^{\frac{b-a}{4 G_N M}} (a b-2 G_N M (a+b))\right)}{2 \left((a-2 G_N M)^2 \sqrt{\frac{b-2 G_N M}{a-2 G_N M}}\right)}\Bigg\} \nonumber \\
     \cosh \left(\frac{t_a-t_b}{4 G_N M}\right)+\pi  \lambda  M (t_a-t_b) \sinh \left(\frac{t_a-t_b}{4 G_N M}\right) &\ll& \Bigg\{\frac{1}{2} e^{\frac{b-a}{4 G_N M}} \sqrt{\frac{b-2 G_N M}{a-2 G_N M}} \nonumber \\
    && -\frac{\lambda  \left(\pi  M (a-b) e^{\frac{b-a}{4 G_N M}} (a b-2 G_N M (a+b))\right)}{2 \left((a-2 G_N M)^2 \sqrt{\frac{b-2 G_N M}{a-2 G_N M}}\right)}\Bigg\} \nonumber
\end{eqnarray}The generalized entropy \( S_{\rm gen} \) can be expressed as  
\begin{eqnarray}
S_{\rm gen} &=& \frac{2 \pi  a^2}{G_N} + \frac{c}{6}\log \left(\frac{256 G_N^4 M^4 \left(4 \pi  b G_N \lambda  M^2+b-2 G_N M\right)^2 e^{\frac{(b-a) \left(4 \pi  G_N \lambda  M^2+1\right)}{2 G_N M}}}{a b \left(4 \pi  G_N \lambda  M^2+1\right)^6}\right) \nonumber \\
&& -\frac{2c}{3} \left(e^{\frac{(a-b) \left(4 \pi  G_N \lambda  M^2+1\right)}{4 G_N M}} \sqrt{\frac{4 \pi  a G_N \lambda  M^2+a-2 G_N M}{4 \pi  b G_N \lambda  M^2+b-2 G_N M}} \cosh \left(\frac{(t_a-t_b) \left(4 \pi  G_N \lambda  M^2+1\right)}{4 G_N M}\right)\right) \ . \nonumber
\end{eqnarray}

Extremizing this expression first with respect to \( t_a \) gives \( t_a = t_b \). Extremizing with respect to \( a \) determines the island's position as:  
\begin{eqnarray}
   a &=& 2 G_N M \left(2 G_N M \left(\frac{e^{-\frac{b}{2 G_N M}-2 \pi  b \lambda  M+1}}{b}-2 \pi  \lambda  M\right)+1\right) \ .
\end{eqnarray}

Including the island’s location, the entanglement entropy becomes:  
\begin{eqnarray}\label{Renyi with island}
    S_{\rm gen} &=& \left(\frac{1}{6} c \log \left(\frac{128 G_N^3 M^3 (b-2 G_N M)^2}{b}\right)+\frac{b}{2 G_N M}+8 \pi  G_N M^2-1\right) \nonumber \\
    && +\frac{2}{3} \pi  \lambda  M \left(\frac{c G_N M (10 G_N M-3 b)}{b-2 G_N M}+3 b-96 \pi  G_N^2 M^3\right) + \mathcal{O}(\lambda^2) \ .
\end{eqnarray}

This result shows that the entropy stabilizes at late times, resolving the entropy paradox.  
\begin{figure}[ht]
    \begin{center}
        \includegraphics[scale=0.75]{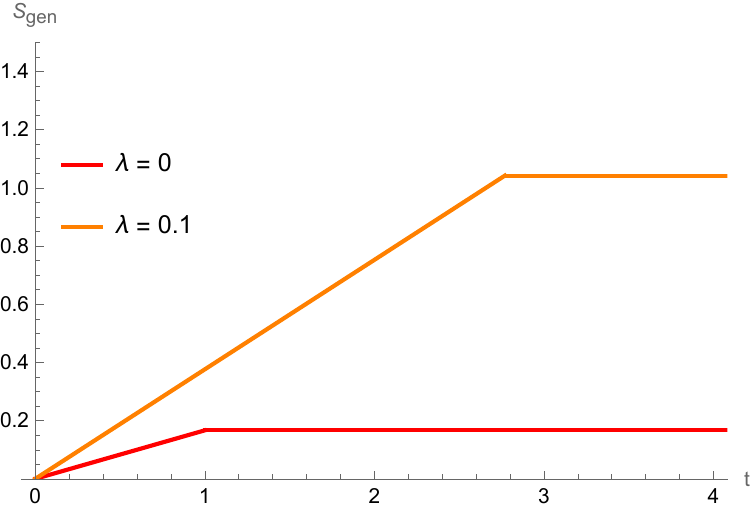}
    \end{center}
    \caption{Plot of \( S_{\rm gen} \) with and without the island.}
    \label{fig:Renyi Sgen}
\end{figure}

In conclusion, comparing results \eqref{S without Island} and \eqref{S with Island} confirms that the island configuration emerges at late times, halting the growth of entanglement entropy. The island boundary is located near the event horizon, contributing directly to the Renyi entropy of the corrected black hole.  

\subsection{Page Time}

Again, the Page time can be determined by comparing the entropies for scenarios without an island \eqref{Renyi without island} and with an island \eqref{Renyi with island}. This critical time corresponds to the intersection of the two entropy curves, marking the transition from a no-island configuration to the formation of an island. After this transition, the entanglement entropy stabilizes at a constant value. The Renyi-corrected Page time is given by 
\begin{equation}
t_{\text{Page}} \sim \frac{96 \pi ^2 G_N^3 M^3}{c}-\frac{1152 \lambda  \left(\pi ^3 G_N^4 M^5\right)}{c}+ \mathcal{O}\left(\lambda ^2\right) \ ,  
\end{equation}  
where \( \lambda \) is the Renyi correction parameter. This result encapsulates the influence of Renyi's corrections and also corrects the Page time of Renyi's corrected case. The Page curve as in Fig.~\ref{fig:Renyi Sgen}, which traces the evolution of radiation entropy, can be constructed by combining the expressions for the entropy in the no-island and island configurations. In the absence of an island, the radiation entropy increases linearly with time. Around the Page time, an island forms near the black hole horizon, leading to the stabilization of entropy. This curve effectively captures the transition and eventual stabilization of entropy due to the island formation.
\section{Tsallis-Cirto's Approach}\label{Sec:Tsallis-Cirto's Approach}

In this section, we investigate the Tsallis-Cirto framework by employing Eq.~\eqref{Tsallis-Cirto Entropy} to compute the Tsallis-Cirto-corrected temperature, derive the corresponding metric function, and analyze the spacetime structure in detail. Starting with Eq.~\eqref{Tsallis-Cirto Entropy}, which characterizes the entropy correction under the Tsallis-Cirto modification, the corrected temperature \( T_{TC} \) is determined as  
\begin{equation}
    T_{TC} = \frac{2^{-2 \delta -3} \pi ^{-\delta -1} M \left(G M^2\right)^{-\delta -1}}{\delta +1} \ .
\end{equation}  
This temperature encapsulates quantum-gravitational effects parameterized by the Tsallis-Cirto parameter \( \delta \). Utilizing this modified temperature, we construct the Tsallis-Cirto-corrected metric function \( f_{TC}(r) \), which reflects deviations from the standard Schwarzschild geometry. For a static, spherically symmetric spacetime, the metric function is expressed as:  
\begin{equation}\label{fTC}
    f_{TC}(r) = 1-\frac{2^{2 \delta +1} \pi ^{\delta } (\delta +1) G M \left(G M^2\right)^{\delta }}{r} \ ,
\end{equation}  
where \( \delta \) denotes the Tsallis-Cirto parameter. By plotting this function, the dependence of the horizon radius on the Tsallis-Cirto parameter can be visualized.  
\begin{figure}[ht]
    \begin{center}
        \includegraphics[scale=.90]{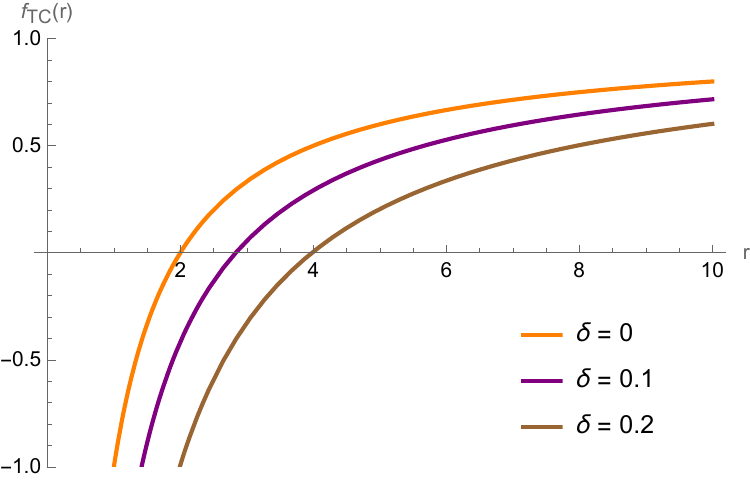} 
    \end{center}
    \caption{Plot of Tsallis-Cirto metric function versus \( r \) for different values of the Tsallis-Cirto parameter.}
    \label{fig:TasalliCirto metric function}
\end{figure}
Again, by using Eq.~\eqref{fTC}, the horizon radius in the Tsallis-Cirto-corrected spacetime is expressed as  
\begin{eqnarray}  
r_h^{{\;}\rm Tsallis-Cirto} &=& 2^{2 \delta +1} \pi ^{\delta } (\delta +1) G M \left(G M^2\right)^{\delta } \nonumber \\
&=& 2 G M+2 \delta  G M \left(\log \left(G M^2\right)+1+2 \log (2)+\log (\pi )\right)+O\left(\delta ^2\right) \ . \nonumber 
\end{eqnarray}  
A direct calculation confirms that the metric function \eqref{fTC} satisfies the Einstein field equations. In addition, the Kretschmann scalar, which serves as an invariant measure of curvature, can be explicitly computed as  
\begin{equation} \label{Tsallis-Cirto Krestmann}  
\mathcal{K} = \frac{3 G \, \pi ^{2 \delta -2} \delta ^2 \left(16 G M^2\right)^{2 \delta -1}}{16\,r^6} \ .  
\end{equation}  
Again similar to the other corrected black holes, the singularity at \( r = r_h^{\;\rm Tsallis-Cirto} \) is a coordinate singularity, and singularity at \( r = 0 \) corresponds to a real curvature singularity, as demonstrated by the behavior of the Kretschmann scalar in Eq.~ \eqref{Tsallis-Cirto Krestmann}. To eliminate the coordinate singularity, we introduce the tortoise coordinate, which can be derived from the metric function, as follows 
\begin{eqnarray}
    r_\star(r) = r + 2^{2 \delta +1} \pi ^{\delta } (\delta +1) G M \left(G M^2\right)^{\delta } \log \left(r-2^{2 \delta +1} \pi ^{\delta } (\delta +1) G M \left(G M^2\right)^{\delta }\right) \ .
\end{eqnarray}  
Similar to Eq.~\eqref{Kruskal Metric} in terms of the Kruskal-like coordinates, it is easy to verify \( \Omega_{TC}(r) \) as the conformal factor incorporating Tsallis-Cirto corrections, given by  
\begin{equation}
    \Omega_{TC}(r) = \frac{\beta}{2 \pi }  e^{-\frac{2 \pi  r_\star(r)}{\beta }} \sqrt{1-\frac{2^{2 \delta +1} \pi ^{\delta } (\delta +1) G M \left(G M^2\right)^{\delta }}{r}} \ .
\end{equation}  
Using the conformal mapping approach, we examine the implications of the Tsallis-Cirto corrections on the entanglement entropy contribution arising from matter fields within this modified spacetime geometry.
\subsection*{Without Island}

During the initial stages of evaporation, when no island forms, the entanglement entropy of the radiation grows indefinitely. Using the Eq.~\eqref{Without Island Entropy} and the Kruskal-like coordinates, the entanglement entropy is given by
\begin{equation}
S_{\rm gen.} = \frac{\beta }{\pi} \cosh \left(\frac{2 \pi  t_b}{\beta }\right) \sqrt{\frac{\left(b-2^{2 \delta +1} \pi ^{\delta } (\delta +1) G M \left(G M^2\right)^{\delta }\right)}{b}} \ ,
\end{equation}  
and in the late-time limit (\( t_b \gg 1 \)), this simplifies to:  
\begin{equation}\label{Tasalli-Crito without island}
S_{\rm gen.} \approx \frac{c t_b}{6}-\frac{c t_b}{6} ( (1+\log (4)+\log (\pi )))\delta \ .
\end{equation}  

This result confirms that, in the absence of an island, the entanglement entropy grows linearly with time, creating a paradox that challenges unitarity and the conservation of information\cite{Nojiri:2022aof}. To resolve this paradox, the island construction is necessary and helps us restore a Page curve consistent with unitarity.
\subsection*{With Island}

For the computation of the entanglement entropy in the presence of an island and in the late time. We Use Kruskal coordinates and Eq.~\eqref{With Island Entropy}, the generalized entropy \( S_{\rm gen} \)\footnote{The approximations we have used here are 
\begin{eqnarray}
    \cosh \left(\frac{(t_a+t_b) M^{-2\delta} }{2^{2 (\delta +1)} (\delta +1)\pi^\delta \left(G_N\right)^{\delta +1}}\right) &\gg& \frac{1}{2} \sqrt{\frac{b-2^{2 \delta +1} \pi ^{\delta } (\delta +1) G_N M \left(G_N M^2\right)^{\delta }}{a-2^{2 \delta +1} \pi ^{\delta } (\delta +1) G_N M \left(G_N M^2\right)^{\delta }} \exp \left(\frac{2^{-2\delta -1}  M (b-a) \left( M^2\right)^{-\delta -1}}{\pi ^{\delta }(\delta +1)G_N^{\delta+1}}\right)}  \nonumber \\
    \cosh \left(\frac{(t_a-t_b) M^{-2\delta} }{2^{2 (\delta +1)} (\delta +1)\pi^\delta \left(G_N\right)^{\delta +1}}\right) &\ll& \frac{1}{2} \sqrt{\frac{b-2^{2 \delta +1} \pi ^{\delta } (\delta +1) G_N M \left(G_N M^2\right)^{\delta }}{a-2^{2 \delta +1} \pi ^{\delta } (\delta +1) G_N M \left(G_N M^2\right)^{\delta }} \exp \left(\frac{2^{-2\delta -1}  M (b-a) \left(M^2\right)^{-\delta -1}}{\pi ^{\delta }(\delta +1)G_N^{\delta+1}}\right)}  \nonumber
\end{eqnarray}} can be  expressed as 
\begin{eqnarray}
S_{\rm gen} &=& \frac{2 \pi  a^2}{G_N}-\Bigg\{\frac{2c}{3} \sqrt{\frac{a-2^{2 \delta +1} \pi ^{\delta } (\delta +1) G_N M \left(G_N M^2\right)^{\delta }}{b-2^{2 \delta +1} \pi ^{\delta } (\delta +1) G_N M \left(G_N M^2\right)^{\delta }}} \exp \left(\frac{2^{-2 \delta -2} \pi ^{-\delta } (a-b) \left(G_N M^2\right)^{-\delta }}{(\delta +1) G_N M}\right)  \nonumber \\
&& \cosh \left(\frac{2^{-2 \delta -2} \pi ^{-\delta } (t_a-t_b) \left(G_N M^2\right)^{-\delta }}{(\delta +1) G_N M}\right) \Bigg\}+\frac{c}{6}  \log \Bigg[\frac{2^{8 \delta +8} \pi ^{4 \delta } (\delta +1)^4 G_N^4 M^4 \left(G_N M^2\right)^{4 \delta } }{a b} \nonumber \\
&& \frac{\left(b-2^{2 \delta +1} \pi ^{\delta } (\delta +1) G_N M \left(G_N M^2\right)^{\delta }\right)^2 \exp \left(\frac{2^{-2 \delta -1} \pi ^{-\delta } (b-a) \left(G_N M^2\right)^{-\delta }}{(\delta +1) G_N M}\right)}{ab}\Bigg] \ .
\end{eqnarray}
Here, We have applied similar approximations to those used in the Barrow case to analyze the late-time behavior. Extremizing this expression with respect to \( t_a \) gives \( t_a = t_b \), and extremizing with respect to \( a \) determines the island location 
\begin{eqnarray}
    a &=& 2^{2 \delta +1} \pi ^{\delta } (\delta +1) G_N M \left(G_N M^2\right)^{\delta } \nonumber \\
&& +\frac{c^2 4^{2 \delta +1} \pi ^{2 \delta } (\delta +1)^2 G_N \left(G_N M^2\right)^{2 \delta +1} \exp \left(1-\frac{b 2^{-2 \delta -1} \pi ^{-\delta } M \left(G_N M^2\right)^{-\delta -1}}{\delta +1}\right)}{\left(b-2^{2 \delta +1} \pi ^{\delta } (\delta +1) G_N M \left(G_N M^2\right)^{\delta }\right) \left(c-3\ 16^{\delta +1} \pi ^{2 \delta +1} (\delta +1)^2 \left(G_N M^2\right)^{2 \delta +1}\right)^2} \ .
\end{eqnarray}

Substituting the island location, the entanglement entropy becomes:  
\begin{eqnarray}\label{Tasalli-Crito with island}
S_{\rm gen} &=& \frac{\delta  \left(\log \left(4 \pi  G_N M^2\right)+1\right) \left(3 b^2-3 b G_N M \left(c+32 \pi  G_N M^2+2\right)+2 G_N^2 M^2 \left(5 c+96 \pi  G_N M^2\right)\right)}{6 G_N M (2 G_N M-b)} \nonumber \\
&&+\left(\frac{1}{6} c \log \left(\frac{128 G_N^3 M^3 (b-2 G_N M)^2}{b}\right)+\frac{b}{2 G_N M}+8 \pi  G_N M^2-1\right) +\mathcal{O}\left(\delta ^2\right) \ .
\end{eqnarray}
This result demonstrates that the entropy stabilizes at late times, resolving the entropy paradox and restoring consistency with unitarity.

\begin{figure}[ht]
    \begin{center}
        \includegraphics[scale=0.75]{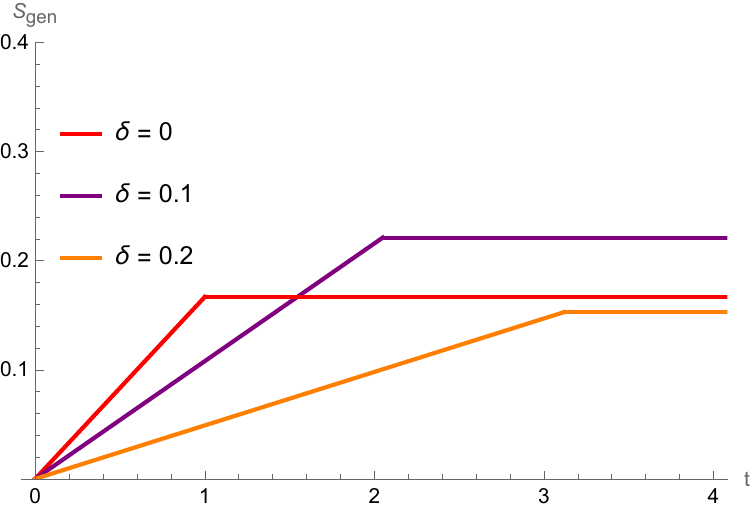}
    \end{center}
    \caption{Plot of \( S_{\rm gen} \) with and without the island.}
    \label{fig:Tasalli-Crito Sgen}
\end{figure}

In conclusion, the comparison of results confirms that the island configuration emerges at late times, arresting the unbounded growth of entanglement entropy\cite{ff, MASI2005217}. The boundary of the island lies near the event horizon and contributes to the Tsallis-Cirto entropy, ensuring consistency with the Page curve and resolving the information paradox.
\subsection*{Page Time}

The Page time can be determined by comparing the entropies i.e.,  without an island \eqref{Tasalli-Crito without island} and with an island \eqref{Tasalli-Crito with island}. This critical time corresponds to the intersection of the two entropy curves, marking the transition from a no-island configuration to the formation of an island. After this transition, the entanglement entropy stabilizes at a constant value. The Tsallis-Cirto-corrected Page time is given by  
\begin{equation}
t_{\text{Page}} \sim \frac{96 \pi ^2 G_N^3 M^3}{c}+\frac{288 \pi ^2   G_N^3 M^3 \left(\log \left(4 \pi  G_N M^2\right)+1\right)}{c} \delta + \mathcal{O}\left(\delta ^2\right) \ ,  
\end{equation}  
where \( \delta \) is the Tsallis-Cirto correction parameter. This result encapsulates the influence of Tsallis-Cirto corrections on the corrected page time in the Tsallis-Crito-corrected spacetime. The Page curve as in Fig.\ref{fig:Tasalli-Crito Sgen}, which shows the evolution of radiation entropy, can be constructed by combining the expressions for the entropy in the no-island and island configurations.



\section{Conclusions}\label{Sec:Conclusion}

This paper investigated the black holes characterized by non-extensive entropy formulations, examining their implications for the information paradox and their resolution using island formulation. The non-extensive correction to Bekenstein-Hawking entropy provides an alternative framework often linked to modifications from the Generalized Uncertainty Principle. Notably, in the limit where the non-extensive parameter vanishes, the entropy recovers its standard Bekenstein-Hawking form. The horizon radius expands in all cases, and restoring classical parameters recovers the Schwarzschild solution. However, altering entropy inevitably impacts other thermodynamic quantities, posing challenges in maintaining consistency within the broader thermodynamic framework. 

\quad We have studied the metric from the non-extensive corrections. It also verified that these solutions satisfied Einstien's Equation, and by computing the Krestchmann Scalar, it is shown that $r=0$ is the true physical singularity of these black holes, too. Furthermore, non-extensive parameter modification and the corrected black holes shift the event horizon from the standard Schwarzschild radius. Then, we explored the Page curves and information paradox resolutions for corrected black holes. In the early stages of evaporation, no island forms as the entanglement entropy is dominated by radiation. At late times, the entropy is determined by the island’s area within the stretched horizon. Using the s-wave approximation and 2D CFT, we computed the entanglement entropy of Hawking radiation and analyzed the effect of non-extensive parameters on the Page curve. Our study is limited to zero or one island case, though more complex formations may exist. Notably, the sharp transition at the Page time could be smoothed by multiple islands. We have also observed that the page time changes with the inclusion of non-extensive parameters and plotted for zero as well as non-zero cases. While our results confirm information conservation in the corrected black hole background, the exact dynamics of information transfer into the radiation zone remain an open question, and we will get back to these issues soon.

\section*{Acknowledgments}
A.S. would like to thank CSIR-HRDG for the financial support received as a Research Associate (RA) working under Project No. CSIR/2024-2025/1107/PHYSICS.



\begin{thebibliography}{99}

\bibitem{Tsallis:2012js}
C.~Tsallis and L.~J.~L.~Cirto,
\emph{Black hole thermodynamical entropy,}
Eur. Phys. J. C \textbf{73} (2013) 2487,
[arXiv:1202.2154 [cond-mat.stat-mech]].

\bibitem{Renyi}
A. ~Renyi,
\emph{On measures of information and entropy,}
Berkeley Symp. on Math. Statist. and Prob. (1961) 547-561.

\bibitem{Tsallis:1987eu}
C.~Tsallis,
\emph{Possible Generalization of Boltzmann-Gibbs Statistics,}
J. Statist. Phys. \textbf{52} (1988) 479-487,

\bibitem{Barrow:2020tzx}
J.~D.~Barrow,
\emph{The Area of a Rough Black Hole,}
Phys. Lett. B \textbf{808} (2020) 135643,
[arXiv:2004.09444 [gr-qc]].

\bibitem{SayahianJahromi:2018irq}
A.~Sayahian Jahromi, S.~A.~Moosavi, H.~Moradpour, J.~P.~Morais Gra\c{c}a, I.~P.~Lobo, I.~G.~Salako and A.~Jawad,
\emph{Generalized entropy formalism and a new holographic dark energy model,}
Phys. Lett. B \textbf{780} (2018) 21-24,
[arXiv:1802.07722 [gr-qc]].

\bibitem{Kaniadakis:2005zk}
G.~Kaniadakis,
\emph{Statistical mechanics in the context of special relativity. II.,}
Phys. Rev. E \textbf{72} (2005) 036108,
[arXiv:cond-mat/0507311 [cond-mat]].

\bibitem{Drepanou:2021jiv}
N.~Drepanou, A.~Lymperis, E.~N.~Saridakis and K.~Yesmakhanova,
\emph{Kaniadakis holographic dark energy and cosmology,}
Eur. Phys. J. C \textbf{82} no.5, (2022) 449,
[arXiv:2109.09181 [gr-qc]].

\bibitem{Nojiri:2021czz}
S.~Nojiri, S.~D.~Odintsov and V.~Faraoni,
\emph{Area-law versus R\'enyi and Tsallis black hole entropies,}
Phys. Rev. D \textbf{104} no.8, (2021) 084030,
[arXiv:2109.05315 [gr-qc]].

\bibitem{Donoghue:1994dn}
J.~F.~Donoghue,
\emph{General relativity as an effective field theory: The leading quantum corrections,}
Phys. Rev. D \textbf{50} (1994) 3874-3888,
[arXiv:gr-qc/9405057 [gr-qc]].

\bibitem{Calmet:2021lny}
X.~Calmet and F.~Kuipers,
\emph{Quantum gravitational corrections to the entropy of a Schwarzschild black hole,}
Phys. Rev. D \textbf{104} no.6, (2021) 66012,
[arXiv:2108.06824 [hep-th]].

\bibitem{Delgado:2022pcc}
R.~C.~Delgado,
\emph{Quantum gravitational corrections to the entropy of a Reissner\textendash{}Nordstr\"om black hole,}
Eur. Phys. J. C \textbf{82} no.3, (2022) 272,
[erratum: Eur. Phys. J. C \textbf{83} (2023) no.6, 468]
[arXiv:2201.08293 [hep-th]].

\bibitem{Wald:1993nt}
R.~M.~Wald,
\emph{Black hole entropy is the Noether charge,}
Phys. Rev. D \textbf{48} no.8, (1993) R3427-R3431,
[arXiv:gr-qc/9307038 [gr-qc]].

\bibitem{Cano:2019ycn}
P.~A.~Cano, S.~Chimento, R.~Linares, T.~Ort\'\i{}n and P.~F.~Ram\'\i{}rez,
\emph{$\alpha'$ corrections of Reissner-Nordstr\"om black holes,}
JHEP \textbf{02} (2020) 031,
[arXiv:1910.14324 [hep-th]].

\bibitem{Yoon:2007aj}
M.~Yoon, J.~Ha and W.~Kim,
\emph{Entropy of Reissner-Nordstrom Black Holes with Minimal Length Revisited,}
Phys. Rev. D \textbf{76} (2007) 047501,
[arXiv:0706.0364 [gr-qc]].

\bibitem{Singh:2023hit}
A.~Singh, P.~Mukherjee and C.~Bhamidipati,
\emph{Thermodynamic curvature of charged black holes with AdS2 horizons,}
Phys. Rev. D \textbf{108}, no.10, 106011 (2023)
[arXiv:2307.11641 [hep-th]].

\bibitem{Akbar:2003mv}
M.~M.~Akbar and S.~Das,
\emph{Entropy corrections for Schwarzschild and Reissner-Nordstr\"om black holes,}
Class. Quant. Grav. \textbf{21} (2004) 1383-1392,
[arXiv:hep-th/0304076 [hep-th]].

\bibitem{Sadeghi:2014zna}
J.~Sadeghi, B.~Pourhassan and F.~Rahimi,
\emph{Logarithmic corrections of charged hairy black holes in (2 + 1) dimensions,}
Can. J. Phys. \textbf{92} no.12, (2014) 1638-1642
[arXiv:1708.07383 [gr-qc]].

\bibitem{Susskind:1994sm} 
  L.~Susskind and J.~Uglum,
  \emph{Black hole entropy in canonical quantum gravity and superstring theory,}
  Phys.\ Rev.\ D {\bf 50}, 2700 (1994)
  doi:10.1103/PhysRevD.50.2700
  [hep-th/9401070].

 
\bibitem{Hawking:1975vcx}
S.~W.~Hawking,
\emph{Particle Creation by Black Holes,}
Commun. Math. Phys. \textbf{43} (1975), 199-220
[erratum: Commun. Math. Phys. \textbf{46} (1976), 206]
doi:10.1007/BF02345020

\bibitem{Hawking:1974rv}
S.~W.~Hawking,
\emph{Black hole explosions,}
Nature \textbf{248} (1974), 30-31
doi:10.1038/248030a0

\bibitem{Hawking:1974sw} 
  S.~W.~Hawking,
  \emph{Particle Creation by Black Holes,}
  Commun.\ Math.\ Phys.\  {\bf 43}, 199 (1975)
  Erratum: [Commun.\ Math.\ Phys.\  {\bf 46}, 206 (1976)].
  

\bibitem{Hawking:1976ra}
S.~W.~Hawking,
\emph{Breakdown of Predictability in Gravitational Collapse,}
Phys. Rev. D \textbf{14} (1976), 2460-2473
doi:10.1103/PhysRevD.14.2460

\bibitem{Maldacena:1997de}
J.~M.~Maldacena, A.~Strominger and E.~Witten,
\emph{Black hole entropy in M theory,}
JHEP \textbf{12} (1997) 002,
[arXiv:hep-th/9711053 [hep-th]].

\bibitem{Solodukhin:2011gn}
S.~N.~Solodukhin,
\emph{Entanglement entropy of black holes,}
Living Rev. Rel. \textbf{14} (2011) 8,
[arXiv:1104.3712 [hep-th]].

\bibitem{Solodukhin:1994yz}
S.~N.~Solodukhin,
\emph{The Conical singularity and quantum corrections to entropy of black hole,}
Phys. Rev. D \textbf{51} (1995) 609-617,
[arXiv:hep-th/9407001 [hep-th]].

\bibitem{Fursaev:1994te}
D.~V.~Fursaev,
\emph{Temperature and entropy of a quantum black hole and conformal anomaly,}
Phys. Rev. D \textbf{51} (1995) 5352-5355,
[arXiv:hep-th/9412161 [hep-th]].

\bibitem{Sen:2012dw}
A.~Sen,
\emph{Logarithmic Corrections to Schwarzschild and Other Non-extremal Black Hole Entropy in Different Dimensions,}
JHEP \textbf{04} (2013) 156,
[arXiv:1205.0971 [hep-th]].



\bibitem{El-Menoufi:2015cqw}
B.~K.~El-Menoufi,
\emph{Quantum gravity of Kerr-Schild spacetimes and the logarithmic correction to Schwarzschild black hole entropy,}
JHEP \textbf{05} (2016) 035,
[arXiv:1511.08816 [hep-th]].


\bibitem{El-Menoufi:2017kew}
B.~K.~El-Menoufi,
\emph{Quantum gravity effects on the thermodynamic stability of 4D Schwarzschild black hole,}
JHEP \textbf{08} (2017) 068,
[arXiv:1703.10178 [gr-qc]].

\bibitem{Cai:2009ua}
R.~G.~Cai, L.~M.~Cao and N.~Ohta,
\emph{Black Holes in Gravity with Conformal Anomaly and Logarithmic Term in Black Hole Entropy,}
JHEP \textbf{04} (2010) 082,
[arXiv:0911.4379 [hep-th]].

\bibitem{Carlip:2000nv}
S.~Carlip,
\emph{Logarithmic corrections to black hole entropy from the Cardy formula,}
Class. Quant. Grav. \textbf{17} (2000)  4175-4186,
[arXiv:gr-qc/0005017 [gr-qc]].


\bibitem{Banerjee:2008cf}
R.~Banerjee and B.~R.~Majhi,
\emph{Quantum Tunneling Beyond Semiclassical Approximation,}
JHEP \textbf{06} (2008) 095,
[arXiv:0805.2220 [hep-th]].

\bibitem{Banerjee:2008fz}
R.~Banerjee and B.~R.~Majhi,
\emph{Quantum Tunneling, Trace Anomaly and Effective Metric,}
Phys. Lett. B \textbf{674} (2009) 218-222,
[arXiv:0808.3688 [hep-th]].

\bibitem{kaul}
R.~K.~Kaul and P.~Majumdar,
\emph{Logarithmic correction to the Bekenstein-Hawking entropy,}
Phys. Rev. Lett. \textbf{84} (2000) 5255-5257,
[arXiv:gr-qc/0002040 [gr-qc]].



\bibitem{Page:1993wv} 
  D.~N.~Page,
  \emph{Information in black hole radiation,}
  Phys.\ Rev.\ Lett.\  {\bf 71}, 3743 (1993)
  [hep-th/9306083].

\bibitem{Page:2013dx} 
  D.~N.~Page,
  \emph{Time Dependence of Hawking Radiation Entropy,}
  JCAP {\bf 1309}, 028 (2013)
  [arXiv:1301.4995 [hep-th]].



\bibitem{Penington:2019npb}
G.~Penington,
\emph{Entanglement Wedge Reconstruction and the Information Paradox,}
[arXiv:1905.08255 [hep-th]].

\bibitem{Almheiri:2019psf}
A.~Almheiri, N.~Engelhardt, D.~Marolf and H.~Maxfield,
\emph{The entropy of bulk quantum fields and the entanglement wedge of an evaporating black hole,}
JHEP \textbf{12} (2019), 063
[arXiv:1905.08762 [hep-th]].



\bibitem{Almheiri:2019hni} 
  A.~Almheiri, R.~Mahajan, J.~Maldacena and Y.~Zhao,
  \emph{The Page curve of Hawking radiation from semiclassical geometry,}
  arXiv:1908.10996 [hep-th].

\bibitem{Almheiri:2019yqk} 
  A.~Almheiri, R.~Mahajan and J.~Maldacena,
\emph{Islands outside the horizon,}
  arXiv:1910.11077 [hep-th].



\bibitem{Ryu:2006bv}
S.~Ryu and T.~Takayanagi,
\emph{Holographic derivation of entanglement entropy from AdS/CFT,}
Phys.\ Rev.\ Lett.\  \textbf{96} (2006), 181602
[arXiv:hep-th/0603001 [hep-th]].

\bibitem{Hubeny:2007xt}
V.~E.~Hubeny, M.~Rangamani and T.~Takayanagi,
\emph{A Covariant holographic entanglement entropy proposal,}
JHEP \textbf{07} (2007), 062
[arXiv:0705.0016 [hep-th]].

\bibitem{Engelhardt:2014gca}
N.~Engelhardt and A.~C.~Wall,
\emph{Quantum Extremal Surfaces: Holographic Entanglement Entropy beyond the Classical Regime,}
JHEP \textbf{01} (2015), 073
[arXiv:1408.3203 [hep-th]].


  
\bibitem{Callan:1994py}
C.~G.~Callan, Jr. and F.~Wilczek,
\emph{On geometric entropy,}
Phys.\ Lett.\ B \textbf{333} (1994), 55-61
[arXiv:hep-th/9401072 [hep-th]].

\bibitem{Holzhey:1994we}
C.~Holzhey, F.~Larsen and F.~Wilczek,
\emph{Geometric and renormalized entropy in conformal field theory,}
Nucl.\ Phys.\ B \textbf{424} (1994), 443-467
[arXiv:hep-th/9403108 [hep-th]].

\bibitem{Calabrese:2009qy}
P.~Calabrese and J.~Cardy,
\emph{Entanglement entropy and conformal field theory,}
J.\ Phys.\ A \textbf{42} (2009), 504005
[arXiv:0905.4013 [cond-mat.stat-mech]].

\bibitem{Penington:2019kki} 
  G.~Penington, S.~H.~Shenker, D.~Stanford and Z.~Yang,
  \emph{Replica wormholes and the black hole interior,}
  arXiv:1911.11977 [hep-th].


\bibitem{Almheiri:2019qdq} 
  A.~Almheiri, T.~Hartman, J.~Maldacena, E.~Shaghoulian and A.~Tajdini,
  \emph{Replica Wormholes and the Entropy of Hawking Radiation,}
  arXiv:1911.12333 [hep-th].




  
\bibitem{Kaniadakis:2002zz}
G.~Kaniadakis,
\emph{Statistical mechanics in the context of special relativity,}
Phys. Rev. E \textbf{66} (2002) 056125.



\bibitem{Czinner:2015eyk}
V.~G.~Czinner and H.~Iguchi,
\emph{R\'enyi Entropy and the Thermodynamic Stability of Black Holes,}
Phys. Lett. B \textbf{752} (2016), 306-310
doi:10.1016/j.physletb.2015.11.061
[arXiv:1511.06963 [gr-qc]].


\bibitem{Nakarachinda:2022gsb}
R.~Nakarachinda, C.~Promsiri, L.~Tannukij and P.~Wongjun,
\emph{Thermodynamics of Black Holes with R\'enyi Entropy from Classical Gravity,}
[arXiv:2211.05989 [gr-qc]].

\bibitem{Volovik:2024}
G.~E.~Volovik, 
\emph{Tsallis-Cirto entropy of black hole and black hole atom,}
arXiv:2409.15362 [hep-th].


\bibitem{Bombelli:1986rw}
L.~Bombelli, R.~K.~Koul, J.~Lee and R.~D.~Sorkin,
\emph{A Quantum Source of Entropy for Black Holes,}
Phys.\ Rev.\ D \textbf{34} (1986), 373-383.

\bibitem{Srednicki:1993im}
M.~Srednicki,
\emph{Entropy and area,}
Phys.\ Rev.\ Lett.\  \textbf{71} (1993), 666-669
[arXiv:hep-th/9303048 [hep-th]].

\bibitem{Faulkner:2013ana}
T.~Faulkner, A.~Lewkowycz and J.~Maldacena,
\emph{Quantum corrections to holographic entanglement entropy,}
JHEP \textbf{11} (2013), 074
[arXiv:1307.2892 [hep-th]].


\bibitem{Dong:2016hjy}
X.~Dong, A.~Lewkowycz and M.~Rangamani,
\emph{Deriving covariant holographic entanglement,}
JHEP \textbf{11} (2016), 028
[arXiv:1607.07506 [hep-th]].

\bibitem{Nutma:2013zea}
T.~Nutma,
\emph{xTras : A field-theory inspired xAct  package for mathematica,}
Comput. Phys. Commun. \textbf{185} (2014) 1719--1738,
[arXiv:1308.3493 [cs.SC]].


\bibitem{DiGennaro:2022grw}
S.~Di Gennaro, H.~Xu and Y.~C.~Ong,
\emph{How barrow entropy modifies gravity: with comments on Tsallis entropy,}
Eur. Phys. J. C \textbf{82} no.11, (2022) 1066,
[arXiv:2207.09271 [gr-qc]].

\bibitem{Anand:2024txo}
A.~Anand and R.~Campos Delgado,
\emph{Modified gravity theories from the Barrow hypothesis,}
EPL \textbf{146} (2024) no.6, 69001
doi:10.1209/0295-5075/ad4c02
[arXiv:2403.13687 [gr-qc]].

\bibitem{Jusufi:2021fek}
K.~ Jusufi et al.,
\emph{Constraints on Barrow Entropy from M87 and S2 Star Observations,}
Universe \textbf{8} no.2, (2022) 102,
[arXiv:2110.07258 [gr-qc]].

\bibitem{Abreu:2024tdv}
E.M.C.~Abreu,
\emph{Surface gravity analysis in Gauss-Bonnet and Barrow black holes,}
[arXiv:2403.02540 [gr-qc]].

\bibitem{Lewkowycz:2013nqa}
A.~Lewkowycz and J.~Maldacena,
\emph{Generalized gravitational entropy,}
JHEP \textbf{08} (2013), 090
[arXiv:1304.4926 [hep-th]].

\bibitem{Nojiri:2022dkr}
S.~Nojiri, S.~D.~Odintsov and T.~Paul,
\emph{Early and late universe holographic cosmology from a new generalized entropy,}
Phys. Lett. B \textbf{831} (2022) 137189,
[arXiv:2205.08876 [gr-qc]].

\bibitem{MASI2005217}  
M.~Masi,  
\emph{A step beyond Tsallis and Renyi entropies},  
Phys. Lett. A \textbf{338} (2005) 217–224,  
doi: https://doi.org/10.1016/j.physleta.2005.01.094{10.1016/j.physleta.2005.01.094}.

\bibitem{Nojiri:2022aof}
S.~Nojiri, S.~D.~Odintsov and V.~Faraoni,
\emph{From nonextensive statistics and black hole entropy to the holographic dark universe,}
Phys. Rev. D \textbf{105} no.4, (2022) 044042,
[arXiv:2201.02424 [gr-qc]].

\bibitem{ff}
E.~Akturk, G.~B.~Bagci, and R.~Sever,  
\emph{Is Sharma-Mittal entropy really a step beyond Tsallis and Renyi entropies?},  
arXiv:cond-mat/0703277 [cond-mat] (2007).



\bibitem{Biro:2011ncf}
T.~S.~Bir\'o and P.~V\'an,
\emph{Zeroth law compatibility of nonadditive thermodynamics,}
Phys. Rev. E \textbf{83}, no.6, 061147 (2011)
doi:10.1103/PhysRevE.83.061147









\end{thebibliography}
\end{document}